# Quantum fluctuations and noise in interferometry and photodetection: Applications in optical sensing and micromanipulation

Masud Mansuripur
Wyant College of Optical Sciences, The University of Arizona, Tucson

**Abstract**. Accurate optical sensing and micromanipulation requires sensitive measurements of the position, orientation, and dynamics of small particles — and sometimes even large objects — under consideration. The signals acquired in the process, including those needed for the feedback control of these particles and objects, are inevitably contaminated by quantum fluctuations and noise that accompany the physical processes of optical interference and photodetection (or photon counting). This paper explores the origins of signal fluctuation and quantum noise that are inevitably associated with such sensitive measurements.

**1. Introduction**. In the classical Maxwell-Lorentz theory of electrodynamics, an electromagnetic (EM) plane-wave propagating in free space has electric field $\boldsymbol{E}(\boldsymbol{r},t) = E_0[\cos(\boldsymbol{k}\cdot\boldsymbol{r} - \omega t + \varphi_0)\boldsymbol{e}' + \sin(\boldsymbol{k}\cdot\boldsymbol{r} - \omega t + \varphi_0)\boldsymbol{e}'']$, where $(\boldsymbol{r},t) = (x,y,z,t)$ is the spacetime coordinates, $\omega$ is the temporal frequency of the field oscillations, $\boldsymbol{k} = (\omega/c)\hat{\boldsymbol{\kappa}}$ is the propagation vector (with $c$ being the speed of light in vacuum, and $\hat{\boldsymbol{\kappa}}$ a unit-vector in the direction of propagation), $E_0$ and $\varphi_0$ are the amplitude and phase of the $E$-field, and $(\boldsymbol{e}', \boldsymbol{e}'')$ is a pair of orthogonal vectors satisfying the relations $\boldsymbol{e}'\cdot\hat{\boldsymbol{\kappa}} = \boldsymbol{e}''\cdot\hat{\boldsymbol{\kappa}} = \boldsymbol{e}'\cdot\boldsymbol{e}'' = 0$ and $\boldsymbol{e}'\cdot\boldsymbol{e}' + \boldsymbol{e}''\cdot\boldsymbol{e}'' = 1$. The parameters $E_0, \varphi_0, \omega, \hat{\boldsymbol{\kappa}}, \boldsymbol{e}', \boldsymbol{e}''$ of the classical theory are measurable entities that can assume arbitrary values subject only to the above restrictions — aside from the trivial constraints $\omega \geq 0$, $E_0 \geq 0$. (The polarization of the $E$-field is conveniently specified by the complex unit-vector $\hat{\boldsymbol{e}} = \boldsymbol{e}' + \mathrm{i}\boldsymbol{e}''$, which satisfies the identity $\hat{\boldsymbol{e}}\cdot\hat{\boldsymbol{e}}^* = 1$.)[1-3]

In contrast, the theory of quantum electrodynamics, which allows $E$-field modes with similarly unrestricted values of $\omega, \hat{\boldsymbol{\kappa}}$, and $\hat{\boldsymbol{e}}$ to exist in free space, tends to impose substantial constraints on the concurrent measurability (as well as acceptable values) of the field's amplitude $E_0$ and phase $\varphi_0$. A single mode $(\omega, \boldsymbol{k}, \hat{\boldsymbol{e}})$ of the quantized EM field can host any superposition of the number-states $|n\rangle$, where $n = 0, 1, 2, \cdots$ is the number of photons in the corresponding state. The energy content of the entire single-mode EM field in the number-state $|n\rangle$ is $(n + \tfrac{1}{2})\hbar\omega$, where $\hbar$ is the reduced Planck constant. In general, the $(\omega, \boldsymbol{k}, \hat{\boldsymbol{e}})$ mode is occupied by a state $|\psi\rangle = \sum_{n=0}^{\infty} c_n |n\rangle$, where the complex coefficient $c_n = |c_n|e^{\mathrm{i}\theta_n}$ is the probability amplitude that the occupying state be $|n\rangle$. What one can or cannot know about $E_0$ and $\varphi_0$ is intimately tied to the distributions of $|c_n|$ and $\theta_n$ over the various number-states $|n\rangle$.[4-8] The main goal of the present paper is to elucidate the relations between the collective of $|c_n|$ and $\theta_n$ on the one hand, and certain measurable features of the EM field on the other. The $E$-field operator $\widehat{\boldsymbol{E}}^{(+)}$ and its conjugate transpose (or adjoint) $\widehat{\boldsymbol{E}}^{(-)}$ introduced in Sec.2 play important roles in the discussions that follow.

Sensitive optical measurements typically involve interferometry and photodetection (or photon counting). While interferometric tests probe and rely upon the phase of the EM field, photodetectors produce a signal that is proportional to the number of photons contained in an EM wavepacket.[2,6,7,9] The Hermitian operator $\widehat{\boldsymbol{E}}^{(-)}\widehat{\boldsymbol{E}}^{(+)}$ for the photon-counting rate yields the expected number of photons in a single-mode wavepacket as $\langle n \rangle = \sum_{n=0}^{\infty} |c_n|^2 n$, and also the expected value of the squared number of photons as $\langle n^2 \rangle = \sum_{n=0}^{\infty} |c_n|^2 n^2$. Fluctuations in the number of counted photons can be lowered by reducing the variance $\langle n^2 \rangle - \langle n \rangle^2$ of the detected number; this can be achieved by narrowing the width of the distribution of $|c_n|$ over the available states $|n\rangle$. In contrast, it is not easy to pinpoint the phase of the $E$-field since, strictly speaking, a phase operator does *not* exist in quantum electro-dynamics.[6] Nevertheless, we will show that it is possible to obtain reasonably accurate measures of the EM phase for an arbitrary state $|\psi\rangle$ of the field. It turns out that a fairly well-defined phase requires a rather broad and uniform distribution of $|c_n|$ — in addition to a linear dependence of the phase $\theta_n$ of the probability amplitude $c_n$ on the photon-number $n$. Clearly, the requirements for low-noise photodetection are incompatible with those for phase accuracy, thus necessitating that certain compromises be made when designing interferometric systems for quantum applications.

In the next section, we describe the $E$-field operators $\hat{\boldsymbol{E}}^{(+)}$ and $\hat{\boldsymbol{E}}^{(-)}$ in the context of a single-mode EM field in free space, then use them to construct operators for photodetection and photon-counting rate. The elementary properties of the quasi-classical (or Glauber) coherent state $|\gamma\rangle$ are discussed in Sec.3. Not only does the coherent state serve as an example that reveals quantum fluctuations as a source of noise in photodetection, but it also plays a pivotal role in setting up heterodyne or homodyne detection schemes needed to monitor the "phase" of any arbitrary state of a single-mode EM field.

The concept of $E$-field quadratures is introduced in Sec.4, where we invoke the homodyne measurement of the components of a classical $E$-field phasor as a segue into the notion of quadrature operators $\hat{E}_p$ and $\hat{E}_q$ for the quantized EM field. The quadrature operators are subsequently defined in Sec.5, where their close association with the quadratures $E_0 \sin \varphi_0$ and $E_0 \cos \varphi_0$ of a classical monochromatic plane-wave with amplitude $E_0$ and phase $\varphi_0$ is established. Quantum fluctuations of the quadrature signals are examined in Sec.6, followed by a derivation in Sec.7 of the Heisenberg uncertainty relation that constrains the product $(\delta E_p)(\delta E_q)$ of the standard deviations of the non-commuting observables $\hat{E}_p$ and $\hat{E}_q$.

Section 8 begins by introducing a special state $|\varphi\rangle$ for single-mode EM fields that exhibits many desirable attributes of a phase eigenstate, namely, a state that has a well-defined, albeit arbitrary phase $\varphi$ in the $[0, 2\pi)$ interval. It turns out, however, that the intriguing properties of $|\varphi\rangle$ are insufficient to render the EM phase an observable entity; the reasons for the non-existence of a Hermitian phase operator in quantum electrodynamics are elaborated in Sec.9.

Additional properties of single-mode EM fields can be studied with the aid of lossless beam-splitters, whose basic properties are reviewed in Sec.10, and also with the aid of optical devices such as the Mach-Zehnder interferometer, which is briefly analyzed in Sec.11. We discuss the degree of first-order coherence of an optical beam in Sec.12, and proceed to examine the degree of second-order coherence in Sec.13. The paper closes with a few concluding remarks in Sec.14.

**2. Photodetection and photon-counting rate**. A single mode of the EM field in free space is typically specified as a plane-wave having temporal frequency $\omega$, propagation vector (or $k$-vector) $\boldsymbol{k} = (\omega/c)\hat{\boldsymbol{\kappa}}$, and polarization state $\hat{\boldsymbol{e}}$. Here, $c = (\mu_0 \varepsilon_0)^{-\frac{1}{2}}$ is the speed of light in vacuum (with $\mu_0$ and $\varepsilon_0$ being the permeability and permittivity of free space), $\hat{\boldsymbol{\kappa}}$ is a real-valued unit-vector aligned with the direction of propagation, and $\hat{\boldsymbol{e}} = \boldsymbol{e}' + \mathrm{i}\boldsymbol{e}''$ is a complex unit-vector satisfying $\hat{\boldsymbol{e}} \cdot \hat{\boldsymbol{\kappa}} = 0$ and $\hat{\boldsymbol{e}} \cdot \hat{\boldsymbol{e}}^* = |\boldsymbol{e}'|^2 + |\boldsymbol{e}''|^2 = 1$.[4-8] In general, $\hat{\boldsymbol{e}}$ with no further restrictions can specify an arbitrary polarization state (i.e., linear, circular, or elliptical) of the EM plane-wave, although imposing an orthogonality constraint on $\boldsymbol{e}'$ and $\boldsymbol{e}''$ (i.e., $\boldsymbol{e}' \cdot \boldsymbol{e}'' = 0$) simplifies any and all calculations without affecting the physics under consideration.

The single-mode EM field is generally occupied by a superposition of the number states $|n\rangle$, where $n = 0, 1, 2, \cdots$ is the number of photons in the corresponding number state. The energy content of the mode in the state $|n\rangle$ is $(n + \frac{1}{2})\hbar\omega$, where $\hbar = 1.054571817 \cdots \times 10^{-34}$ Joule·second is Planck's reduced constant. Thus, the mode's ground state $|0\rangle$ (i.e., vacuum) has energy $\frac{1}{2}\hbar\omega$, which is then incremented by $\hbar\omega$ for each added photon. The general state of a single mode of the EM field is $|\psi\rangle = \sum_{n=0}^{\infty} c_n |n\rangle$, with the complex number $c_n$ being the probability amplitude of $|n\rangle$; as always, the normalization condition is $\sum_{n=0}^{\infty} |c_n|^2 = 1$.

In the Heisenberg picture, where the spacetime dependence is embedded in the operators, the single-mode electric field operator $\widehat{\boldsymbol{E}}(\boldsymbol{r},t)$ is expressed in terms of the annihilation and creation operators $\hat{a}$ and $\hat{a}^\dagger$. Denoting by $V$ the free space volume occupied by the $(\omega, \boldsymbol{k}, \hat{\boldsymbol{e}})$ mode, we have[7] in SI units → $\widehat{\boldsymbol{E}}(\boldsymbol{r},t) = \mathrm{i}\sqrt{\hbar\omega/(2\varepsilon_0 V)}\,\{\hat{\boldsymbol{e}}\exp[\mathrm{i}(\boldsymbol{k}\cdot\boldsymbol{r} - \omega t)]\,\hat{a} - \hat{\boldsymbol{e}}^*\exp[-\mathrm{i}(\boldsymbol{k}\cdot\boldsymbol{r} - \omega t)]\,\hat{a}^\dagger\}.$ (1)[†]

The first term appearing on the right-hand side of Eq.(1) is commonly referred to as $\widehat{\boldsymbol{E}}^{(+)}$, and the second term, which is the Hermitian conjugate of $\widehat{\boldsymbol{E}}^{(+)}$, is known as $\widehat{\boldsymbol{E}}^{(-)}$. Consequently,

$$\widehat{\boldsymbol{E}}(\boldsymbol{r},t) = \widehat{\boldsymbol{E}}^{(+)}(\boldsymbol{r},t) + \widehat{\boldsymbol{E}}^{(-)}(\boldsymbol{r},t). \qquad (2)$$

The annihilation operator $\hat{a}$ acting on the number state $|n\rangle$ brings out the square root of $n$ while reducing the occupation number by 1; that is, $\hat{a}|n\rangle = \sqrt{n}|n-1\rangle$. Consequently, the $E$-field "amplitude" emerging from $\widehat{\boldsymbol{E}}^{(+)}|n\rangle$ turns out to be $\mathrm{i}\sqrt{n\hbar\omega/(2\varepsilon_0 V)}\,\hat{\boldsymbol{e}}$, whose associated energy-density is readily found by analogy with the classical $E$-field energy-density of $\varepsilon_0 E_0^2$.[‡] In the case of an EM plane-wave that carries $n$ photons within a spatial volume $V$, this $E$-field energy-density equals one-half the total energy-density, namely, $n\hbar\omega/(2V)$, with the remaining half of the energy-density residing in the magnetic field.

The creation operator $\hat{a}^\dagger$ acting on the number state $|n\rangle$ brings out the square root of $n+1$ while raising the occupation number by 1; that is, $\hat{a}^\dagger|n\rangle = \sqrt{n+1}|n+1\rangle$. Thus, $\hat{a}^\dagger\hat{a}|n\rangle = n|n\rangle$ and $\hat{a}\hat{a}^\dagger|n\rangle = (n+1)|n\rangle$, which lead to the commutation relation $[\hat{a},\hat{a}^\dagger] = \hat{a}\hat{a}^\dagger - \hat{a}^\dagger\hat{a} = \hat{\mathbb{1}}$. The Hamiltonian for the single-mode EM field (a harmonic oscillator) is $\widehat{\mathcal{H}} = \hbar\omega(\hat{a}^\dagger\hat{a} + \tfrac{1}{2}\hat{\mathbb{1}})$. The action of the Hamiltonian on the number state $|n\rangle$ yields $\widehat{\mathcal{H}}|n\rangle = (n + \tfrac{1}{2})\hbar\omega|n\rangle$, thus revealing the total energy content of the mode to be that of $n$ photons, each of energy $\hbar\omega$, plus the ground-state (or vacuum) energy $\tfrac{1}{2}\hbar\omega$ associated with our single-mode field.

In experiments involving photodetection (or photon counting) the proper operator for the EM energy flux is the Hermitian operator $(\hbar\omega c/V)\hat{a}^\dagger\hat{a}$, which extracts the photon number $n$ from the state $|n\rangle$, multiplies it by $\hbar\omega/V$ to obtain the energy per unit volume of the mode, then multiplies it again by the speed of light $c$ to arrive at the rate of flow of energy per unit area per unit time. This energy flux operator may be written equivalently as $2Z_0^{-1}\widehat{\boldsymbol{E}}^{(-)}(\boldsymbol{r},t)\cdot\widehat{\boldsymbol{E}}^{(+)}(\boldsymbol{r},t)$, with $Z_0 = (\mu_0/\varepsilon_0)^{1/2}$ being the impedance of free space. Note that, upon division by $\hbar\omega$, the time-averaged energy flux operator $2Z_0^{-1}\widehat{\boldsymbol{E}}^{(-)}\cdot\widehat{\boldsymbol{E}}^{(+)}$ yields the operator for the photon-counting rate (per unit area per unit time). Also, in contrast to the Hamiltonian $\widehat{\mathcal{H}}$, the energy flux operator does *not* take into account the contribution $\tfrac{1}{2}\hbar\omega$ of the vacuum energy, which is appropriate, considering that conventional photodetectors cannot capture the energy of the underlying free space.

In the case of double photon-counting at $(\boldsymbol{r}_1, t_1)$ and $(\boldsymbol{r}_2, t_2)$, one must use an intensity operator that takes account of the various $E$-field components at both locations 1 and 2. Considering that

$$I_1 I_2 = (E_{x1}^2 + E_{y1}^2 + E_{z1}^2)(E_{x2}^2 + E_{y2}^2 + E_{z2}^2) = \sum_{i,j=x,y,z} E_{i,1}^2 E_{j,2}^2 = \sum_{i,j=x,y,z} E_{i,1}^* E_{j,2}^* E_{j,2} E_{i,1}, \quad (3)$$

the correct operator is found to be proportional to $\sum_{i,j=x,y,z} \hat{E}_{i,1}^{(-)} \hat{E}_{j,2}^{(-)} \hat{E}_{j,2}^{(+)} \hat{E}_{i,1}^{(+)}$. Here, the ordering of the operators is of crucial significance, since annihilation of one photon at location 1 reduces the probability of detecting another at location 2, and vice-versa. Thus, the annihilation operators $\hat{E}_{i,1}^{(+)}$ (at location 1) and $\hat{E}_{j,2}^{(+)}$ (at location 2) must act on the state $|\psi\rangle$ of the EM field prior to the action of the corresponding creation operators $\hat{E}_{j,2}^{(-)}$ and $\hat{E}_{i,1}^{(-)}$.

---

[†] The carets over $\boldsymbol{\kappa}$, $\boldsymbol{e}$, and $\boldsymbol{e}^*$ identify these symbols as unit vectors, whereas those over $a$, $a^\dagger$, and $\boldsymbol{E}$ signify that these symbols represent quantum-mechanical operators.

[‡] A classical electric field $\boldsymbol{E}(\boldsymbol{r},t) = E_0[\hat{\boldsymbol{e}}\,e^{\mathrm{i}(\boldsymbol{k}\cdot\boldsymbol{r}-\omega t)} + \hat{\boldsymbol{e}}^*e^{-\mathrm{i}(\boldsymbol{k}\cdot\boldsymbol{r}-\omega t)}] = 2E_0[\boldsymbol{e}'\cos(\boldsymbol{k}\cdot\boldsymbol{r} - \omega t) - \boldsymbol{e}''\sin(\boldsymbol{k}\cdot\boldsymbol{r} - \omega t)]$ has energy-density $\tfrac{1}{2}\varepsilon_0|\boldsymbol{E}(\boldsymbol{r},t)|^2 = 2\varepsilon_0 E_0^2[(\boldsymbol{e}'\cdot\boldsymbol{e}')\cos^2(\boldsymbol{k}\cdot\boldsymbol{r} - \omega t) + (\boldsymbol{e}''\cdot\boldsymbol{e}'')\sin^2(\boldsymbol{k}\cdot\boldsymbol{r} - \omega t)]$, which, upon time-averaging, becomes $2\varepsilon_0 E_0^2(\tfrac{1}{2}|\boldsymbol{e}'|^2 + \tfrac{1}{2}|\boldsymbol{e}''|^2) = \varepsilon_0 E_0^2$.

**3. On the quasi-classical (or Glauber) coherent state**. Before proceeding further, we must explore the elementary properties of a special state of the single-mode propagating EM field in free space. The quasi-classical (or Glauber) coherent state of the $(\omega, \boldsymbol{k}, \hat{\boldsymbol{e}})$ mode is a special superposition of all the number states $|n\rangle$, defined as $|\gamma\rangle = e^{-|\gamma|^2/2} \sum_{n=0}^{\infty} \gamma^n |n\rangle / \sqrt{n!}$, with the complex number $\gamma$ being the characteristic parameter of the state.[4] The coherent state is an eigenstate of the annihilation operator with $\gamma$ as its eigenvalue; that is, $\hat{a}|\gamma\rangle = \gamma|\gamma\rangle$. The conjugate transpose of this identity is $\langle \gamma|\hat{a}^\dagger = \langle \gamma|\gamma^*$. (Recalling that $\hat{a}$ is not Hermitian, it should not be surprising that its eigenvalues are complex numbers.) Applying the photon-number operator $\hat{n} = \hat{a}^\dagger \hat{a}$, which is Hermitian, to the coherent state $|\gamma\rangle$ yields the expected (or average) number of photons in the state as

$$\langle n \rangle = \langle \gamma|\hat{n}|\gamma\rangle = \langle \gamma|\hat{a}^\dagger \hat{a}|\gamma\rangle = \gamma^* \gamma \langle \gamma|\gamma\rangle = |\gamma|^2. \quad (4)$$

Similarly, the photon-number variance of a coherent state is found to be

$$\text{var}(n) = \langle n^2 \rangle - \langle n \rangle^2 = \langle \gamma|\hat{a}^\dagger \hat{a}\hat{a}^\dagger \hat{a}|\gamma\rangle - |\gamma|^4 = \langle \gamma|\hat{a}^\dagger(\hat{a}^\dagger \hat{a} + 1)\hat{a}|\gamma\rangle - |\gamma|^4$$

$$= \gamma^{*2} \gamma^2 \langle \gamma|\gamma\rangle + |\gamma|^2 - |\gamma|^4 = |\gamma|^2. \quad (5)$$

The action of the energy flux operator $2Z_0^{-1} \hat{\boldsymbol{E}}^{(-)} \cdot \hat{\boldsymbol{E}}^{(+)}$ on the coherent state $|\gamma\rangle$ yields

$$\langle \gamma|2Z_0^{-1} \hat{\boldsymbol{E}}^{(-)}(\boldsymbol{r},t) \cdot \hat{\boldsymbol{E}}^{(+)}(\boldsymbol{r},t)|\gamma\rangle = (\hbar\omega c/V)\langle \gamma|\hat{a}^\dagger \hat{a}|\gamma\rangle = (\hbar\omega c/V)|\gamma|^2. \quad (6)$$

Clearly, this is the time-averaged rate of flow of energy (per unit area per unit time) for our single-mode coherent state with the characteristic parameter $\gamma$, harboring an average number $|\gamma|^2$ of photons within the confines of its (vast) spatial volume $V$.

Let us mention in passing that $|\gamma\rangle$ is *not* an eigenstate of the Hamiltonian $\hat{\mathcal{H}} = \hbar\omega(\hat{a}^\dagger \hat{a} + \frac{1}{2})$. Consequently, in the Schrödinger picture, the temporal variation of $|\gamma\rangle$ differs from that of typical stationary states.[1,2] Considering that each number-state $|n\rangle$ has energy $(n + \frac{1}{2})\hbar\omega$, which gives it a time-dependent phase-factor $e^{-\mathrm{i}(n+\frac{1}{2})\omega t}$, the coherent state's time-dependence can be expressed as

$$|\gamma(t)\rangle = e^{-\mathrm{i}\omega t/2} e^{-|\gamma|^2/2} \sum_{n=0}^{\infty} (\gamma e^{-\mathrm{i}\omega t})^n |n\rangle / \sqrt{n!}. \quad (7)$$

Thus, according to this equation—and aside from the inconsequential phase-factor $e^{-\mathrm{i}\omega t/2}$—the time-dependence of the coherent state is embedded within its characteristic parameter $\gamma e^{-\mathrm{i}\omega t}$. Needless to say, this temporal variation is relevant only in the Schrödinger picture. In contrast, the Heisenberg picture relegates all temporal variations to the operators while treating the state of the system as time-independent.

Another interesting fact about coherent states is that two such states, say, $|\gamma_1\rangle$ and $|\gamma_2\rangle$ with $\gamma_1 \neq \gamma_2$, are *not* mutually orthogonal. This is readily demonstrated, as follows:

$$\langle \gamma_1|\gamma_2\rangle = e^{-\frac{1}{2}(|\gamma_1|^2 + |\gamma_2|^2)} \sum_{m=0}^{\infty} \sum_{n=0}^{\infty} \langle m|\gamma_1^{*m} \gamma_2^n |n\rangle / \sqrt{m! n!}$$

$$= e^{-\frac{1}{2}(|\gamma_1|^2+|\gamma_2|^2)} \sum_{m=0}^{\infty} (\gamma_1^* \gamma_2)^m/m! = e^{\gamma_1^* \gamma_2 - \frac{1}{2}(|\gamma_1|^2+|\gamma_2|^2)} \neq 0. \tag{8}$$

The above identity leads to $|\langle\gamma_1|\gamma_2\rangle|^2 = e^{\gamma_1^*\gamma_2 + \gamma_1 \gamma_2^* - |\gamma_1|^2 - |\gamma_2|^2} = e^{-|\gamma_1 - \gamma_2|^2}$, indicating that the overlap between $|\gamma_1\rangle$ and $|\gamma_2\rangle$ would be small if their parameters $\gamma_1$ and $\gamma_2$ were far apart when viewed in the complex plane.

The coherent state $|\gamma\rangle$ can be generated by the action of the operator $\hat{\Gamma}(\gamma) = e^{-|\gamma|^2/2} e^{\gamma \hat{a}^\dagger}$ on the vacuum state $|0\rangle$, as follows:

$$\hat{\Gamma}(\gamma)|0\rangle = e^{-|\gamma|^2/2} e^{\gamma \hat{a}^\dagger}|0\rangle = e^{-|\gamma|^2/2} \sum_{n=0}^{\infty}[(\gamma \hat{a}^\dagger)^n/n!]\,|0\rangle = e^{-|\gamma|^2/2} \sum_{n=0}^{\infty} \gamma^n |n\rangle/\sqrt{n!} = |\gamma\rangle. \tag{9}$$

This, however, is not the only way to generate a coherent state. There exists a unitary operator $\hat{T}(\gamma)$, known as the translation operator, that also generates the coherent state $|\gamma\rangle$ upon acting on the vacuum state $|0\rangle$.[5] To describe the translation operator, we need an important operator identity, which we set out to prove in the next couple of paragraphs.

We begin by proving the operator identity $e^{\hat{A}} e^{\hat{B}} = e^{\hat{A}+\hat{B}+\frac{1}{2}[\hat{A},\hat{B}]}$, which is valid when both $\hat{A}$ and $\hat{B}$ happen to commute with $[\hat{A},\hat{B}]$. (This result is a special case of the so-called Campbell-Baker-Hausdorff formula.[2]) The proof requires the following identities:

$$\hat{A}\hat{B} = \hat{B}\hat{A} + [\hat{A},\hat{B}],$$

$$\hat{A}^2 \hat{B} = \hat{A}\hat{B}\hat{A} + \hat{A}[\hat{A},\hat{B}] = (\hat{B}\hat{A}^2 + [\hat{A},\hat{B}]\hat{A}) + [\hat{A},\hat{B}]\hat{A} = \hat{B}\hat{A}^2 + 2[\hat{A},\hat{B}]\hat{A},$$

$$\hat{A}^3 \hat{B} = \hat{A}\hat{B}\hat{A}^2 + 2\hat{A}[\hat{A},\hat{B}]\hat{A} = (\hat{B}\hat{A}^3 + [\hat{A},\hat{B}]\hat{A}^2) + 2[\hat{A},\hat{B}]\hat{A}^2 = \hat{B}\hat{A}^3 + 3[\hat{A},\hat{B}]\hat{A}^2,$$

$$\vdots$$

$$\hat{A}^n \hat{B} = \hat{B}\hat{A}^n + n[\hat{A},\hat{B}]\hat{A}^{n-1}. \tag{10}$$

Next, we define the operator $\hat{f}(s) = e^{s\hat{A}} e^{s\hat{B}}$ for an arbitrary parameter $s$, and prove the identity

$$e^{s\hat{A}} \hat{B} = \sum_{n=0}^{\infty}(s^n \hat{A}^n/n!)\hat{B} = \sum_{n=0}^{\infty} s^n (\hat{A}^n \hat{B})/n!$$

$$= \hat{B}\sum_{n=0}^{\infty}(s^n \hat{A}^n/n!) + [\hat{A},\hat{B}]\sum_{n=1}^{\infty}[s^n \hat{A}^{n-1}/(n-1)!] = \hat{B}e^{s\hat{A}} + [\hat{A},\hat{B}]s e^{s\hat{A}}. \tag{11}$$

The derivative with respect to $s$ of $\hat{f}(s)$ is readily evaluated with the aid of Eq.(11), as follows:

$$\frac{d}{ds}\hat{f}(s) = \hat{A}e^{s\hat{A}}e^{s\hat{B}} + e^{s\hat{A}}\hat{B}e^{s\hat{B}} = \hat{A}e^{s\hat{A}}e^{s\hat{B}} + \hat{B}e^{s\hat{A}}e^{s\hat{B}} + [\hat{A},\hat{B}]s e^{s\hat{A}}e^{s\hat{B}}$$

$$= (\hat{A} + \hat{B} + [\hat{A},\hat{B}]s)\hat{f}(s). \tag{12}$$

Considering that $\hat{f}(0) = \hat{\mathbb{1}}$, the solution of the above differential equation is seen to be

$$\hat{f}(s) = e^{(\hat{A}+\hat{B})s + \frac{1}{2}[\hat{A},\hat{B}]s^2}. \tag{13}$$

Setting $s = 1$, we arrive at the desired identity. Note that the commutativity of $\hat{B}$ and $[\hat{A},\hat{B}]$ was not needed in the above derivation. It thus appears that the validity of $e^{\hat{A}} e^{\hat{B}} = e^{\hat{A}+\hat{B}+\frac{1}{2}[\hat{A},\hat{B}]}$ requires that either $\hat{A}$ or $\hat{B}$ (but not necessarily both) commute with $[\hat{A},\hat{B}]$.

The translation operator $\hat{T}(\gamma) = e^{\gamma \hat{a}^\dagger - \gamma^* \hat{a}}$, where $\gamma$ is an arbitrary complex number while $\hat{a}$ and $\hat{a}^\dagger$ are the standard annihilation and creation operators satisfying $[\hat{a},\hat{a}^\dagger] = \hat{\mathbb{1}}$, may now be written as $\hat{T}(\gamma) = e^{-\frac{1}{2}|\gamma|^2} e^{\gamma \hat{a}^\dagger} e^{-\gamma^* \hat{a}}$, considering that $[\gamma \hat{a}^\dagger, -\gamma^* \hat{a}] = \gamma \gamma^*[\hat{a},\hat{a}^\dagger] = |\gamma|^2 \hat{\mathbb{1}}$. Note that both $\gamma\hat{a}^\dagger$ and $-\gamma^*\hat{a}$ commute with the constant operator $|\gamma|^2\hat{\mathbb{1}}$. One can easily prove that $\hat{T}(\gamma)$ is a unitary operator, since it satisfies $\hat{T}(\gamma)\hat{T}^\dagger(\gamma) = \hat{T}^\dagger(\gamma)\hat{T}(\gamma) = \hat{\mathbb{1}}$, as follows:

$$\hat{T}(\gamma)\hat{T}^\dagger(\gamma) = (e^{\gamma\hat{a}^\dagger - \gamma^*\hat{a}})(e^{\gamma^*\hat{a} - \gamma\hat{a}^\dagger}) = e^{\frac{1}{2}[(\gamma\hat{a}^\dagger - \gamma^*\hat{a}),(\gamma^*\hat{a} - \gamma\hat{a}^\dagger)]} = e^0 = \hat{\mathbb{1}}. \qquad (14)$$

The unitary translation operator $\hat{T}(\gamma)$ is a generator of the quasi-classical (i.e., Glauber) coherent state $|\gamma\rangle$, since

$$\hat{T}(\gamma)|0\rangle = e^{-\frac{1}{2}|\gamma|^2} e^{\gamma\hat{a}^\dagger} e^{-\gamma^*\hat{a}} |0\rangle = e^{-\frac{1}{2}|\gamma|^2} [\textstyle\sum_{n=0}^{\infty} \gamma^n (\hat{a}^\dagger)^n / n!]\, |0\rangle$$

$$= e^{-\frac{1}{2}|\gamma|^2} \textstyle\sum_{n=0}^{\infty} (\gamma^n / \sqrt{n!}) |n\rangle = |\gamma\rangle. \qquad (15)$$

Application of $\hat{T}^\dagger(\gamma)$, the inverse of $\hat{T}(\gamma)$, to both sides of Eq.(15) now yields $\hat{T}^\dagger(\gamma)|\gamma\rangle = |0\rangle$.

---

**Digression**: The translation operator gets its name from the following property:

$$\hat{T}^\dagger(\gamma)\hat{a}\hat{T}(\gamma) = (e^{\frac{1}{2}|\gamma|^2} e^{\gamma^*\hat{a}}\, e^{-\gamma\hat{a}^\dagger})\hat{a}(e^{-\frac{1}{2}|\gamma|^2} e^{\gamma\hat{a}^\dagger} e^{-\gamma^*\hat{a}})$$

$$\boxed{\text{see Eq.(11)}} \to \quad = e^{\gamma^*\hat{a}}(e^{-\gamma\hat{a}^\dagger}\hat{a})(e^{\gamma\hat{a}^\dagger} e^{-\gamma^*\hat{a}}) = e^{\gamma^*\hat{a}}\{(\hat{a} - \gamma[\hat{a}^\dagger,\hat{a}])e^{-\gamma\hat{a}^\dagger}\}(e^{\gamma\hat{a}^\dagger} e^{-\gamma^*\hat{a}})$$

$$\boxed{\text{see Eq.(11)}} \to \quad = (\hat{a} + \gamma)(e^{\gamma^*\hat{a}} e^{-\gamma\hat{a}^\dagger})(e^{\gamma\hat{a}^\dagger} e^{-\gamma^*\hat{a}}) = (\hat{a} + \gamma)\hat{T}^\dagger(\gamma)\hat{T}(\gamma) = \hat{a} + \gamma. \qquad (16)$$

The conjugate transpose of the above equation is $\hat{T}^\dagger(\gamma)\hat{a}^\dagger\hat{T}(\gamma) = \hat{a}^\dagger + \gamma^*$.

---

**4. Monitoring optical phase via homodyne (or heterodyne) technique: $E$-field quadratures**. A classical plane-wave of frequency $\omega_1$, propagating in free space along the $k$-vector $\boldsymbol{k}_1 = (\omega_1/c)\hat{\boldsymbol{\kappa}}$ and having linear polarization along the unit-vector $\hat{\boldsymbol{e}}$ with a complex amplitude $E_1 = |E_1|e^{i\varphi_1}$, has the following (real-valued) electric field:

$$\boldsymbol{E}_1(\boldsymbol{r},t) = \mathrm{Re}\big[E_1 e^{i(\boldsymbol{k}_1 \cdot \boldsymbol{r} - \omega_1 t)}\hat{\boldsymbol{e}}\big] = |E_1|\cos(\boldsymbol{k}_1 \cdot \boldsymbol{r} - \omega_1 t + \varphi_1)\,\hat{\boldsymbol{e}} \qquad \text{real-valued unit-vector}$$

$$= [|E_1|\cos(\varphi_1)\cos(\boldsymbol{k}_1 \cdot \boldsymbol{r} - \omega_1 t) - |E_1|\sin(\varphi_1)\sin(\boldsymbol{k}_1 \cdot \boldsymbol{r} - \omega_1 t)]\hat{\boldsymbol{e}}. \qquad (17)$$

The coefficients $|E_1|\cos\varphi_1$ and $|E_1|\sin\varphi_1$ appearing on the right-hand side of Eq.(17) are the so-called quadratures of the $E$-field,[7] which, as shown in Fig.1, correspond to the real and imaginary components of the phasor representing the complex $E$-field at $\boldsymbol{r} = 0$ and $t = 0$.

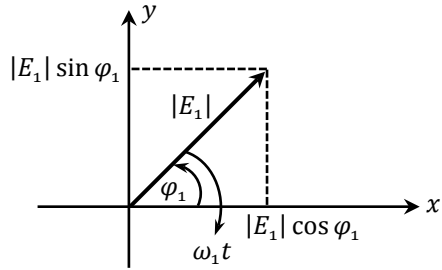

**Fig.1**. A phasor having the components $|E_1|\cos\varphi_1$ and $|E_1|\sin\varphi_1$ at $(\boldsymbol{r},t) = (0,0)$ describes the temporal evolution of the linearly-polarized plane-wave whose $E$-field is given in Eq.(17). As the phasor rotates clockwise within the complex $xy$-plane at the constant angular velocity $\omega_1$, its projection onto the $x$-axis (i.e., the real axis) emulates the time-dependence of the $E$-field at $\boldsymbol{r} = 0$.

The time-averaged Poynting vector of the plane-wave is $|E_1|^2\hat{\boldsymbol{\kappa}}/(2Z_0)$, with $Z_0$ being the impedance of free space. Thus, the rate of arrival of electromagnetic energy (per unit area per unit time) at the surface of a detector (placed perpendicular to $\widehat{\boldsymbol{\kappa}}$) is $|E_1|^2/(2Z_0)$. Assuming the rate of photoelectron generation by the detector is proportional to its rate of capture of the arriving photons (energy $=\hbar\omega_1$), the detector's photocurrent in response to the incident plane-wave will be proportional to $|E_1|^2/(2Z_0\hbar\omega_1)$. Note that the $E$-field magnitude $|E_1|$ defined via Eq.(17) is twice as large as a similar entity that might be associated with the $\widehat{\boldsymbol{E}}^{(\pm)}$ operators emerging from Eq.(1), hence the consistency of the photon-counting rate defined here with that given in Sec.2.

In heterodyne detection, an incoming signal with frequency $\omega_1$ and electric-field amplitude $E_1$ is mixed with a local oscillator having frequency $\omega_2$ and $E$-field amplitude $E_2$, as depicted in Fig.2. (A homodyne detection system is similar, except for the frequency $\omega_2$ of the local oscillator being the same as the source frequency; that is, $\omega_2 = \omega_1$.) An adjustable retroreflector is used in the system of Fig.2 to impart a controlled phase $\varphi_2$ to the local oscillator's $E$-field, which, at the point of entrance to the beam-splitter, is written as $E_2 = |E_2|e^{i\varphi_2}$. The 50/50 beam-splitter, having Fresnel reflection and transmission coefficients $(\rho,\tau) = (1/\sqrt{2}, i/\sqrt{2})$, subsequently divides the two beams between its output ports and proceeds to combine them, so that the $E$-field arriving at detector 1 is

$$\text{Re}\big[|E_1|e^{-i(\omega_1 t-\varphi_1)} + |E_2|e^{-i(\omega_2 t-\varphi_2-\frac{1}{2}\pi)}\big]/\sqrt{2}, \tag{18a}$$

while the $E$-field reaching detector 2 is

$$\text{Re}\big[|E_1|e^{-i(\omega_1 t-\varphi_1-\frac{1}{2}\pi)} + |E_2|e^{-i(\omega_2 t-\varphi_2)}\big]/\sqrt{2}. \tag{18b}$$

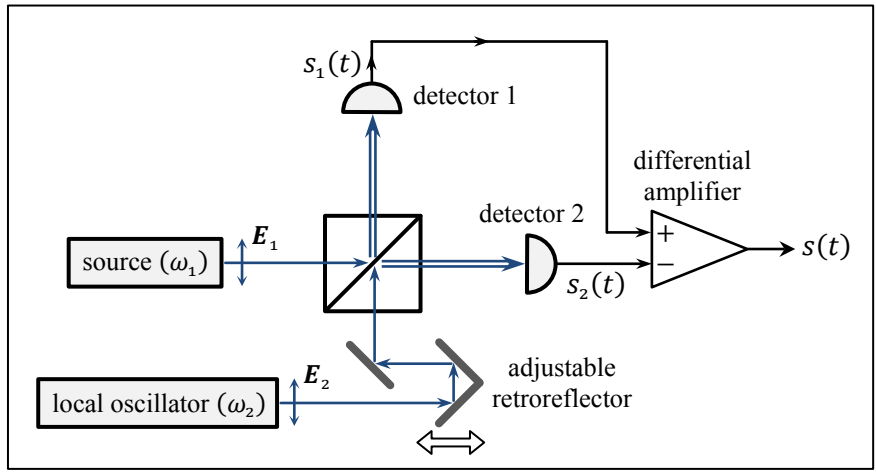

**Fig.2**. In heterodyne detection, an incoming signal with frequency $\omega_1$ and $E$-field amplitude $E_1 = |E_1|e^{i\varphi_1}$ is mixed with a local oscillator signal having frequency $\omega_2$ and $E$-field amplitude $E_2$. An adjustable retro-reflector introduces a controlled phase $\varphi_2$ into the local oscillator's $E$-field, while a 50/50 beam-splitter divides the incoming beams between its output ports. The detected signals are fed to a differential amplifier, whose output $s(t)$ is the difference between the electric signals $s_1(t)$ and $s_2(t)$ emerging from the two detectors. (A homodyne detection system is similar except for the frequency $\omega_2$ of the local oscillator being the same as the source frequency; that is, $\omega_2 = \omega_1$.)

The detected signals are fed to a balanced differential amplifier, whose output signal $s(t)$ is the difference between the electric signals $s_1(t)$ and $s_2(t)$ that emerge from the two detectors. Ignoring the proportionality coefficient between each detector's signal and the corresponding incident $E$-field intensity, we will have

$$s_1(t) = \tfrac{1}{2}[|E_1|\cos(\omega_1 t - \varphi_1) + |E_2|\sin(\omega_2 t - \varphi_2)]^2, \tag{19a}$$

$$s_2(t) = \tfrac{1}{2}[|E_1|\sin(\omega_1 t - \varphi_1) + |E_2|\cos(\omega_2 t - \varphi_2)]^2. \tag{19b}$$

Consequently,

$$s(t) = s_1(t) - s_2(t) = \tfrac{1}{2}|E_1|^2 \cos[2(\omega_1 t - \varphi_1)] - \tfrac{1}{2}|E_2|^2 \cos[2(\omega_2 t - \varphi_2)]$$

$$+ |E_1||E_2|\sin[(\omega_2 - \omega_1)t - \varphi_2 + \varphi_1]. \tag{20}$$

At optical frequencies, the high-frequency terms (i.e., those oscillating at $2\omega_1$ and $2\omega_2$) appearing in $s(t)$ are automatically eliminated, since conventional photodetectors cannot respond to such rapid oscillations. At lower frequencies (e.g., microwaves), these signals must be removed via low-pass filtering. In any event, the filtered output $s(t)$ of the differential amplifier emerges at the difference frequency $\omega_2 - \omega_1$, with a phase $\varphi_1 - \varphi_2$ and an amplitude proportional to $|E_1||E_2|$. At $t = 0$, the output signal will be proportional to $|E_1|\sin\varphi_1$ if the adjustable delay is set such that $\varphi_2 = 0$, and to $|E_1|\cos\varphi_1$ if $\varphi_2$ is set to $-\pi/2$. In this way, the quadratures of the source waveform $E_1(r,t)$ of Eq.(17) are measured. Note that, according to Eq.(20), a strong local oscillator substantially amplifies the output signal in proportion to $|E_2|$.

In the case of homodyne detection, where $\omega_2 = \omega_1$, the (filtered) output $s(t)$ of the differential amplifier is time-independent, being directly proportional to the quadrature $|E_1|\sin\varphi_1$ when $\varphi_2 = 0$, and to $|E_1|\cos\varphi_1$ when $\varphi_2 = -\pi/2$. The sum of the photocurrents of the two detectors is now seen from Eqs.(19) to be proportional to $(|E_1|^2 + |E_2|^2)/(2Z_0\hbar\omega_1)$, where the rapidly oscillating terms at frequency $2\omega_1$ have been eliminated. (One may invoke the identities $\cos^2 x = \tfrac{1}{2}(1 + \cos 2x)$ and $\sin^2 x = \tfrac{1}{2}(1 - \cos 2x)$ to observe that the individual $E$-field intensities arriving at the detectors continue to harbor the rapidly oscillating terms.) In contrast, the difference of the two photocurrents appearing in Eq.(20) should scale as $|E_1||E_2|\sin(\varphi_1 - \varphi_2)/(Z_0\hbar\omega_1)$ — i.e., *without* a factor of 2 in the denominator — since it is *not* affected by the time-averaging of the incident intensities that is generally required for the removal of the high-frequency terms.

**5. Quantum operators for *E*-field quadratures**. In quantum optics, one uses homodyne detection to extract the quadratures of a single-mode $(\omega_1, \mathbf{k}_1, \hat{\mathbf{e}}_1)$ beam in the state $|\psi_1\rangle$ by mixing it with a quasi-classical coherent beam $|\gamma_2\rangle$ of the same frequency $\omega_2 = \omega_1$ and the same linear polarization $\hat{\mathbf{e}}_2 = \hat{\mathbf{e}}_1$. The joint state of the two independent beams arriving at the input ports 1 and 2 of the beam-splitter depicted in the system of Fig.2 may thus be denoted by $|\psi_1, \gamma_2\rangle$. The annihilation operators at the exit ports of the splitter — corresponding to the superposed beams that reach detectors 1 and 2 — are now given by [2,5-7]

$$\text{For the beam arriving at detector 1:} \quad \hat{a} = \rho\hat{a}_1 + \tau\hat{a}_2 = (\hat{a}_1 + i\hat{a}_2)/\sqrt{2}, \tag{21a}$$

$$\text{For the beam arriving at detector 2:} \quad \hat{a} = \tau\hat{a}_1 + \rho\hat{a}_2 = (i\hat{a}_1 + \hat{a}_2)/\sqrt{2}. \tag{21b}$$

The photodetection rate at detector 1 is governed by the following operator:

$$\hat{E}^{(-)}\hat{E}^{(+)}|_{D_1} = (\hbar\omega_1/4\varepsilon_0 V)(\hat{a}_1^\dagger - i\hat{a}_2^\dagger)(\hat{a}_1 + i\hat{a}_2) = (\hbar\omega/4\varepsilon_0 V)(\hat{a}_1^\dagger\hat{a}_1 + \hat{a}_2^\dagger\hat{a}_2 + i\hat{a}_1^\dagger\hat{a}_2 - i\hat{a}_2^\dagger\hat{a}_1). \tag{22}$$

Consequently, aside from a proportionality coefficient, the expected value of the signal $s_1$ emerging from detector 1 is given by

$$\langle s_1 \rangle = \langle \psi_1, \gamma_2 | \hat{E}^{(-)}\hat{E}^{(+)} | \psi_1, \gamma_2 \rangle$$

$$= (\hbar\omega/4\varepsilon_0 V)[\langle\psi_1|\hat{a}_1^\dagger\hat{a}_1|\psi_1\rangle + \langle\gamma_2|\hat{a}_2^\dagger\hat{a}_2|\gamma_2\rangle + i\gamma_2\langle\psi_1|\hat{a}_1^\dagger|\psi_1\rangle - i\gamma_2^*\langle\psi_1|\hat{a}_1|\psi_1\rangle]. \tag{23}$$

Similarly, the photodetection-rate operator and the expected signal $\langle s_2 \rangle$ at detector 2 are found to be

$$\hat{E}^{(-)}\hat{E}^{(+)}|_{D_2} = (\hbar\omega/4\varepsilon_0 V)(-i\hat{a}_1^\dagger + \hat{a}_2^\dagger)(i\hat{a}_1 + \hat{a}_2) = (\hbar\omega/4\varepsilon_0 V)(\hat{a}_1^\dagger\hat{a}_1 + \hat{a}_2^\dagger\hat{a}_2 - i\hat{a}_1^\dagger\hat{a}_2 + i\hat{a}_2^\dagger\hat{a}_1). \quad (24)$$

$$\langle s_2\rangle = \langle\psi_1,\gamma_2|\hat{E}^{(-)}\hat{E}^{(+)}|\psi_1,\gamma_2\rangle$$
$$= (\hbar\omega/4\varepsilon_0 V)\big[\langle\psi_1|\hat{a}_1^\dagger\hat{a}_1|\psi_1\rangle + \langle\gamma_2|\hat{a}_2^\dagger\hat{a}_2|\gamma_2\rangle - i\gamma_2\langle\psi_1|\hat{a}_1^\dagger|\psi_1\rangle + i\gamma_2^*\langle\psi_1|\hat{a}_1|\psi_1\rangle\big]. \quad (25)$$

The differential signal at the amplifier output is finally obtained by subtracting $s_2$ from $s_1$, yielding

$$\langle s\rangle = \langle s_1 - s_2\rangle = (\hbar\omega/2\varepsilon_0 V)\big[i|\gamma_2|e^{i\varphi_2}\langle\psi_1|\hat{a}_1^\dagger|\psi_1\rangle - i|\gamma_2|e^{-i\varphi_2}\langle\psi_1|\hat{a}_1|\psi_1\rangle\big]$$
$$= -(\hbar\omega/2\varepsilon_0 V)|\gamma_2|\big[\sin(\varphi_2)\langle\psi_1|(\hat{a}_1 + \hat{a}_1^\dagger)|\psi_1\rangle + \cos(\varphi_2)\langle\psi_1|i(\hat{a}_1 - \hat{a}_1^\dagger)|\psi_1\rangle\big]. \quad (26)$$

The quadrature operators are obtained from the above equation as $\hat{E}_p = i\sqrt{\hbar\omega/2\varepsilon_0 V}\,(\hat{a}_1^\dagger - \hat{a}_1)$ and $\hat{E}_q = \sqrt{\hbar\omega/2\varepsilon_0 V}\,(\hat{a}_1^\dagger + \hat{a}_1)$ by setting $\varphi_2 = 0$ and $\varphi_2 = -\pi/2$, respectively. It is seen that, in the special case of $|\psi_1\rangle = |\gamma_1\rangle$, where the source beam itself is quasi-classical (i.e., coherent), $\langle\psi_1|\hat{E}_p|\psi_1\rangle$ equals $\sqrt{2\hbar\omega/\varepsilon_0 V}\,|\gamma_1|\sin(\varphi_{\gamma_1})$ and $\langle\psi_1|\hat{E}_q|\psi_1\rangle$ equals $\sqrt{2\hbar\omega/\varepsilon_0 V}\,|\gamma_1|\cos(\varphi_{\gamma_1})$, as anticipated. As for the remaining coefficient on the right-hand side of Eq.(26), by now it is reduced to $\sqrt{\hbar\omega/2\varepsilon_0 V}\,|\gamma_2|$, which is only one-half of the local oscillator's $E$-field magnitude $|E_2|$. However, including the prefactor $2Z_0^{-1}$ of the $\hat{E}^{(-)}\hat{E}^{(+)}$ operator in Eqs.(22)-(25), brings Eq.(26) into complete alignment with Eq.(20), which has been derived in Sec.4 for classical EM fields. Similarly, the sum signal $\langle s_1 + s_2\rangle$, obtained from Eqs.(23) and (25), when multiplied by $2Z_0^{-1}$, yields the overall rate of flow of EM energy (per unit area per unit time), in complete accord with the classical result obtained in Sec.4.

**6. Measurement noise accompanying quadrature signals.** An estimate of the fluctuations of the differential signal $s(t)$ can be found by evaluating the variance $\langle s^2\rangle - \langle s\rangle^2$ of the amplifier's output.[7] The average (or expected value) $\langle s\rangle$ of the homodyne signal is already available in Eq.(26). To obtain $\langle s^2\rangle$, one must evaluate the expected value of the quantum operator corresponding to the classical expression $s^2 = (s_1 - s_2)^2$, namely,

$$\langle\psi_1,\gamma_2|\big(\underbrace{\hat{E}^{(-)}\hat{E}^{(+)}|_{D_1}}_{\text{see Eq.(22)}} - \underbrace{\hat{E}^{(-)}\hat{E}^{(+)}|_{D_2}}_{\text{see Eq.(24)}}\big)^2|\psi_1,\gamma_2\rangle = (\hbar\omega/2\varepsilon_0 V)^2\langle\psi_1,\gamma_2|(i\hat{a}_1^\dagger\hat{a}_2 - i\hat{a}_2^\dagger\hat{a}_1)^2|\psi_1,\gamma_2\rangle$$

$$= (\hbar\omega/2\varepsilon_0 V)^2\langle\psi_1,\gamma_2|(-\hat{a}_1^\dagger\hat{a}_2\hat{a}_1^\dagger\hat{a}_2 + \hat{a}_1^\dagger\hat{a}_2\hat{a}_2^\dagger\hat{a}_1 + \hat{a}_2^\dagger\hat{a}_1\hat{a}_1^\dagger\hat{a}_2 - \hat{a}_2^\dagger\hat{a}_1\hat{a}_2^\dagger\hat{a}_1)|\psi_1,\gamma_2\rangle \quad \boxed{\hat{a}\hat{a}^\dagger = \hat{a}^\dagger\hat{a} + \hat{\mathbb{1}}}$$

$$= (\hbar\omega/2\varepsilon_0 V)^2\big[-\gamma_2^2\langle\psi_1|\hat{a}_1^\dagger\hat{a}_1^\dagger|\psi_1\rangle + (1+|\gamma_2|^2)\langle\psi_1|\hat{a}_1^\dagger\hat{a}_1|\psi_1\rangle + |\gamma_2|^2\langle\psi_1|\hat{a}_1\hat{a}_1^\dagger|\psi_1\rangle - \gamma_2^{*2}\langle\psi_1|\hat{a}_1\hat{a}_1|\psi_1\rangle\big]$$

$$= (\hbar\omega/2\varepsilon_0 V)^2\big\{\langle\psi_1|\hat{a}_1^\dagger\hat{a}_1|\psi_1\rangle - |\gamma_2|^2\big[(\cos\varphi_2 + i\sin\varphi_2)^2\langle\psi_1|\hat{a}_1^\dagger\hat{a}_1^\dagger|\psi_1\rangle - \langle\psi_1|\hat{a}_1^\dagger\hat{a}_1 + \hat{a}_1\hat{a}_1^\dagger|\psi_1\rangle$$
$$+(\cos\varphi_2 - i\sin\varphi_2)^2\langle\psi_1|\hat{a}_1\hat{a}_1|\psi_1\rangle\big]\big\}$$

$$= (\hbar\omega/2\varepsilon_0 V)^2\big\{\langle\psi_1|\hat{a}_1^\dagger\hat{a}_1|\psi_1\rangle - |\gamma_2|^2\big[(\cos^2\varphi_2 - \sin^2\varphi_2)\langle\psi_1|\hat{a}_1^\dagger\hat{a}_1^\dagger + \hat{a}_1\hat{a}_1|\psi_1\rangle - \langle\psi_1|\hat{a}_1^\dagger\hat{a}_1 + \hat{a}_1\hat{a}_1^\dagger|\psi_1\rangle$$
$$+ i\sin(2\varphi_2)\langle\psi_1|(\hat{a}_1^\dagger\hat{a}_1^\dagger - \hat{a}_1\hat{a}_1)|\psi_1\rangle\big]\big\}$$

$$= (\hbar\omega/2\varepsilon_0 V)^2\big\{\langle\psi_1|\hat{a}_1^\dagger\hat{a}_1|\psi_1\rangle - |\gamma_2|^2\big[\cos^2(\varphi_2)\langle\psi_1|(\hat{a}_1 - \hat{a}_1^\dagger)^2|\psi_1\rangle - \sin^2(\varphi_2)\langle\psi_1|(\hat{a}_1 + \hat{a}_1^\dagger)^2|\psi_1\rangle$$
$$\underbrace{\phantom{}}_{\text{½ local oscillator's }E\text{-field magnitude}} - \sin(2\varphi_2)\langle\psi_1|i(\hat{a}_1\hat{a}_1 - \hat{a}_1^\dagger\hat{a}_1^\dagger)|\psi_1\rangle\big]\big\}$$

$$= \big(\sqrt{\hbar\omega/2\varepsilon_0 V}\,|\gamma_2|\big)^2\big[\sin^2(\varphi_2)\langle\psi_1|\hat{E}_q^2|\psi_1\rangle + \cos^2(\varphi_2)\langle\psi_1|\hat{E}_p^2|\psi_1\rangle + \sin(2\varphi_2)\langle\psi_1|i(\hat{a}_1\hat{a}_1 - \hat{a}_1^\dagger\hat{a}_1^\dagger)|\psi_1\rangle\big]$$
$$+ (\hbar\omega/2\varepsilon_0 V)^2\langle\psi_1|\hat{a}_1^\dagger\hat{a}_1|\psi_1\rangle. \quad (27)$$

The last term on the right-hand side of the above equation is due to vacuum fluctuations that enter through port 2 of the beam-splitter, then mix with the source signal that arrives at port 1 — this may be appreciated by setting $\gamma_2 = 0$ and noting that, while $\langle s \rangle$ given by Eq.(26) vanishes, the $\langle s^2 \rangle$ of Eq.(27) continues to have a nonzero value.[§] For sufficiently large values of $|\gamma_2|$, the contribution of the last term of Eq.(27) to the photon-counting fluctuations of $s(t)$ can be ignored.

If we now set $\varphi_2 = -\pi/2$, the only surviving term within the square brackets on the right-hand side of Eq.(27) will be $\langle \psi_1 | \hat{E}_q^2 | \psi_1 \rangle$, which, after subtracting $\langle \psi_1 | \hat{E}_q | \psi_1 \rangle^2$, yields the variance of the quadrature corresponding to $|E_1| \cos \varphi_1$. Similarly, by setting $\varphi_2 = 0$ and noting that the surviving term within the square brackets of Eq.(27) is $\langle \psi_1 | \hat{E}_p^2 | \psi_1 \rangle$, one may proceed to evaluate the variance of the quadrature corresponding to $|E_1| \sin \varphi_1$ by subtracting $\langle \psi_1 | \hat{E}_p | \psi_1 \rangle^2$.

**7. Quadrature observables and the Heisenberg uncertainty principle**. Having defined the quadrature operators $\hat{E}_q = \sqrt{\hbar \omega / (2 \varepsilon_0 V)} \, (\hat{a}^\dagger + \hat{a})$ and $\hat{E}_p = \mathrm{i} \sqrt{\hbar \omega / (2 \varepsilon_0 V)} \, (\hat{a}^\dagger - \hat{a})$, we may now express the single-mode $E$-field operator of Eq.(1) in the following alternative form:

$$\hat{\boldsymbol{E}}(\boldsymbol{r}, t) = \mathrm{i} \sqrt{\hbar \omega / (2 \varepsilon_0 V)} \, \{ \hat{\boldsymbol{e}} \exp[\mathrm{i}(\boldsymbol{k} \cdot \boldsymbol{r} - \omega t)] \, \hat{a} - \hat{\boldsymbol{e}}^* \exp[-\mathrm{i}(\boldsymbol{k} \cdot \boldsymbol{r} - \omega t)] \, \hat{a}^\dagger \}$$

$$= [\sin(\boldsymbol{k} \cdot \boldsymbol{r} - \omega t) \hat{E}_p - \cos(\boldsymbol{k} \cdot \boldsymbol{r} - \omega t) \hat{E}_q] \boldsymbol{e}''$$

$$- [\cos(\boldsymbol{k} \cdot \boldsymbol{r} - \omega t) \hat{E}_p + \sin(\boldsymbol{k} \cdot \boldsymbol{r} - \omega t) \hat{E}_q] \boldsymbol{e}'. \qquad (28)$$

It is seen that, for the real and imaginary components $\boldsymbol{e}'$ and $\boldsymbol{e}''$ of the polarization unit-vector $\hat{\boldsymbol{e}}$, the corresponding $E$-field amplitudes assume the general form of classical phasors. Indeed, for a quasi-classical (or coherent) beam, the expected value of the $\boldsymbol{e}'$ component of the above $E$-field matches the classical $E$-field of Eq.(17) after a 90° phase shift; the same is true of the $\boldsymbol{e}''$ component, albeit after a 180° phase shift. Considering that $\hat{E}_q$ and $\hat{E}_p$ are Hermitian operators whose commutator is given by

$$[\hat{E}_q, \hat{E}_p] = \mathrm{i}(\hbar \omega / \varepsilon_0 V)[\hat{a}, \hat{a}^\dagger] = \mathrm{i}(\hbar \omega / \varepsilon_0 V), \qquad (29)$$

one can invoke the Heisenberg uncertainty relation to deduce that $(\delta E_q)(\delta E_p) \geq \hbar \omega / (2 \varepsilon_0 V)$. Here, the variance (or dispersion) of the quadrature $E_q$ is given by $(\delta E_q)^2 = \langle \psi | \hat{E}_q^2 | \psi \rangle - \langle \psi | \hat{E}_q | \psi \rangle^2$; a similar expression holds for $\delta E_p$. The quadrature components $\hat{E}_p$ and $\hat{E}_q$ of the single-mode $E$-field are observables (say, via homodyne detection), thus enabling one to visualize the behavior of the sinusoidally varying $E$-field as a function of time (at a fixed point $\boldsymbol{r}_0$ in space) by means of a rotating phasor such as that depicted in Fig.1. In the $xy$-plane of the phasor, the $x$-component is $\langle \hat{E}_q \rangle + \delta E_q$ and the $y$-component is $\langle \hat{E}_p \rangle + \delta E_p$, with the random nature of each component's standard deviation (i.e., uncertainty around the mean value) properly incorporated into the diagram. For a linearly polarized, single-mode plane-wave of frequency $\omega$, the magnitude $E(\boldsymbol{r}_0, t)$ for each realization of the $E$-field is a sinusoidal function of time (with frequency $\omega$), whose phase and amplitude are determined by the orientation (at $t = 0$) and the length of the corresponding phasor. Note that the random features of the phasor cannot violate the constraint $(\delta E_q)(\delta E_p) \geq \hbar \omega / (2 \varepsilon_0 V)$ imposed by the uncertainty principle.[7,10] In the special case of the Glauber coherent state $|\gamma\rangle$, it is rather straightforward to show that the quadrature observables satisfy $\delta E_p = \delta E_q = (\hbar \omega / 2 \varepsilon_0 V)^{1/2}$.

---

[§] Setting $|\psi_1\rangle = |0\rangle$ has a similar effect on the output signal of the differential amplifier in that, while $\langle s \rangle$ vanishes, $\langle s^2 \rangle$ acquires the value of $(\hbar \omega / 2 \varepsilon_0 V)^2 |\gamma_2|^2$ due to mixing of the local oscillator signal with the vacuum fluctuations that enter through port 1 of the beam-splitter.

**8. Phase "eigenstates" for single-mode electromagnetic waves in free space.**[6,11] Consider a single-mode EM field $(\omega, \boldsymbol{k}, \hat{\boldsymbol{e}})$ occupied by a superposition of all the number states $|n\rangle$, each with a probability amplitude $e^{in\varphi}/\sqrt{2\pi}$, where the arbitrary phase angle $\varphi$ belongs to $[0, 2\pi)$; that is,

$$|\varphi\rangle = (2\pi)^{-\frac{1}{2}} \sum_{n=0}^{\infty} e^{in\varphi} |n\rangle, \qquad (0 \leq \varphi < 2\pi). \tag{30}$$

The probability of a given state $|\psi\rangle = \sum_{n=0}^{\infty} c_n |n\rangle$ of the $(\omega, \boldsymbol{k}, \hat{\boldsymbol{e}})$ mode to be in the $|\varphi\rangle$ state (i.e., to have the specific phase value $\varphi$), is found by projecting $|\psi\rangle$ onto $|\varphi\rangle$, then squaring the absolute value of the corresponding probability amplitude; in other words,

$$P_\psi(\varphi) = |\langle \varphi | \psi \rangle|^2 = (2\pi)^{-1} |\sum_{n=0}^{\infty} c_n e^{-in\varphi}|^2 = (2\pi)^{-1} \sum_{m=0}^{\infty} \sum_{n=0}^{\infty} c_m c_n^* e^{i(n-m)\varphi}. \tag{31}$$

Integrating the above expression over $\varphi$ (from 0 to $2\pi$) reduces the double-sum to a single sum, which is then readily evaluated, yielding $\sum_{n=0}^{\infty} |c_n|^2 = 1.0$ for the total area under the function. The probability distribution $P_\psi(\varphi)$ is thus found to be properly normalized, as it should.

In the case of $|\psi\rangle = |n\rangle$, Eq.(31) yields $P_\psi(\varphi) = (2\pi)^{-1}$, indicating that all phase angles between 0 and $2\pi$ are equally likely to reside within the number state $|n\rangle$. Another way of reaching the same conclusion is by expressing the number state $|n\rangle$ as a superposition of all phase states $|\varphi\rangle$, each with an amplitude of $e^{-in\varphi}/\sqrt{2\pi}$, as follows:

$$\int_{\varphi=0}^{2\pi} (e^{-in\varphi}/\sqrt{2\pi}) |\varphi\rangle \mathrm{d}\varphi = (2\pi)^{-1} \int_{\varphi=0}^{2\pi} e^{-in\varphi} \sum_{m=0}^{\infty} e^{im\varphi} |m\rangle \, \mathrm{d}\varphi$$

$$= (2\pi)^{-1} \sum_{m=0}^{\infty} [\int_{\varphi=0}^{2\pi} e^{i(m-n)\varphi} \mathrm{d}\varphi] |m\rangle = |n\rangle. \tag{32}$$

In contrast, for the Glauber coherent state $|\gamma\rangle = e^{-\frac{1}{2}|\gamma|^2} \sum_{n=0}^{\infty} (\gamma^n/\sqrt{n!}) |n\rangle$, with $\gamma = |\gamma| e^{i\theta}$, we have

$$P_\gamma(\varphi) = |\langle \varphi | \gamma \rangle|^2 = (2\pi)^{-1} e^{-|\gamma|^2} |\sum_{n=0}^{\infty} (|\gamma|^n e^{in(\theta - \varphi)}/\sqrt{n!})|^2$$

$$= (2\pi)^{-1} e^{-|\gamma|^2} \sum_{m=0}^{\infty} \sum_{n=0}^{\infty} |\gamma|^{m+n} e^{i(m-n)(\theta - \varphi)}/\sqrt{m! \, n!}. \tag{33}$$

If one chooses to define the phase eigenstates $|\varphi\rangle$ of Eq.(30) over the interval $[\theta - \pi, \theta + \pi)$ — instead of the conventional $[0, 2\pi)$ — then $P_\gamma(\varphi)$ of Eq.(33) will exhibit even symmetry around its central point $\varphi = \theta$, in which case the expected value of $\varphi$ turns out to be $\langle \varphi \rangle = \theta$. The variance of $\varphi$ can then be computed by evaluating $\int_{\theta-\pi}^{\theta+\pi} (\varphi - \theta)^2 P_\gamma(\varphi) \mathrm{d}\varphi$ (via integration by parts), as follows:

$$\mathrm{var}(\varphi) = \langle \varphi^2 \rangle - \langle \varphi \rangle^2 = \tfrac{1}{3}\pi^2 + 2e^{-|\gamma|^2} \underset{n \neq m}{\sum_{m=0}^{\infty} \sum_{n=0}^{\infty}} (-1)^{m-n} |\gamma|^{m+n}/[(m-n)^2 \sqrt{m! \, n!}]$$

$$= \tfrac{1}{3}\pi^2 + 4e^{-|\gamma|^2} \sum_{k=1}^{\infty} \sum_{n=0}^{\lfloor (k-1)/2 \rfloor} (-|\gamma|)^k / [(k-2n)^2 \sqrt{(k-n)! \, n!}]. \tag{34}$$

When the above expression is numerically evaluated, the variance of $\varphi$ is found to drop (monotonically and rapidly) with an increasing average number of photons, $\langle n \rangle = |\gamma|^2$, that reside in the coherent mode. While in the vacuum state (i.e., $\gamma = 0$), we have $\mathrm{var}(\varphi) = \tfrac{1}{3}\pi^2$, by the time $\langle n \rangle$ reaches 5, the variance of $\varphi$ has essentially dropped to zero.[6]

The set of all phase eigenstates $|\varphi\rangle$, with $\varphi$ being a phase angle within a continuous $2\pi$ interval, forms a complete orthonormal basis for the single-mode field $(\omega, \boldsymbol{k}, \hat{\boldsymbol{e}})$.[6] One way to ascertain the completeness of the set is as follows:

$$\int_{\varphi=0}^{2\pi} |\varphi\rangle\langle\varphi| \mathrm{d}\varphi = (2\pi)^{-1} \int_{\varphi=0}^{2\pi} \sum_{m=0}^{\infty} \sum_{n=0}^{\infty} e^{i(m-n)\varphi} |m\rangle\langle n| \mathrm{d}\varphi = \sum_{n=0}^{\infty} |n\rangle\langle n| = \hat{\mathbb{1}}. \tag{35}$$

To confirm the orthonormality of the phase eigenstates, we write

$$\langle\varphi'|\varphi\rangle = (2\pi)^{-1} \sum_{m=0}^{\infty}\sum_{n=0}^{\infty} e^{\mathrm{i}(n\varphi-m\varphi')}\langle m|n\rangle = (2\pi)^{-1} \lim_{N\to\infty}\sum_{n=0}^{N} e^{\mathrm{i}n(\varphi-\varphi')}$$

$$= (2\pi)^{-1} \lim_{N\to\infty} [1 - e^{\mathrm{i}(N+1)(\varphi-\varphi')}]/[1 - e^{\mathrm{i}(\varphi-\varphi')}] \quad \leftarrow \boxed{\sum_{n=0}^{N} x^n = (1 - x^{N+1})/(1-x)}$$

$$= (2\pi)^{-1} \lim_{N\to\infty} \left\{ \frac{1 - \cos[(N+1)(\varphi-\varphi')] - \mathrm{i}\sin[(N+1)(\varphi-\varphi')]}{-2\mathrm{i}e^{\mathrm{i}(\varphi-\varphi')/2} \sin[(\varphi-\varphi')/2]} \right\} \quad \leftarrow \boxed{\cos\theta = 1 - 2\sin^2(\theta/2)}$$

$$= \frac{1}{2\pi e^{\mathrm{i}(\varphi-\varphi')/2}} \lim_{N\to\infty} \left\{ \frac{\sin[(N+1)(\varphi-\varphi')]}{2\sin[(\varphi-\varphi')/2]} + \mathrm{i}\frac{\sin^2[(N+1)(\varphi-\varphi')/2]}{\sin[(\varphi-\varphi')/2]} \right\}. \tag{36}$$

The denominators in the preceding expression will be well approximated by $(\varphi - \varphi')/2$ when $\varphi'$ happens to be close to $\varphi$, in which case the first term inside the curly brackets can be written as

$$\boxed{\mathrm{sinc}(x) = \sin(\pi x)/(\pi x)} \to (N + 1)\mathrm{sinc}[(N + 1)(\varphi - \varphi')/\pi], \quad \leftarrow \boxed{\text{area under this function is } \pi} \tag{37a}$$

while the second term becomes

$$\mathrm{i}(N + 1)\mathrm{sinc}[(N + 1)(\varphi - \varphi')/2\pi] \times \sin[(N + 1)(\varphi - \varphi')/2]. \tag{37b}$$

However, as soon as $\varphi'$ deviates a little from $\varphi$, both ratios of the sine functions appearing in Eq.(36) become negligible compared to the peak value of each function at $\varphi = \varphi'$. Thus, the first term inside the curly brackets approaches $\pi\delta(\varphi - \varphi')$, whereas the second term, being an odd function of $\varphi - \varphi'$, becomes null. All in all, the inner product $\langle\varphi'|\varphi\rangle$ converges to $\tfrac{1}{2}\delta(\varphi - \varphi')$, thus confirming the orthonormality of the set of all phase eigenstates.[**]

A useful property of the phase eigenstates $|\varphi\rangle$ is that, when acted upon by $e^{\mathrm{i}\varphi_0\hat{n}}$, they shift to $|\varphi + \varphi_0\rangle$; here $\varphi_0$ is an arbitrary phase and $\hat{n} = \hat{a}^\dagger\hat{a}$ is the photon-number operator. In other words,

$$e^{\mathrm{i}\varphi_0\hat{n}}|\varphi\rangle = (2\pi)^{-1}\sum_{n=0}^{\infty} e^{\mathrm{i}n\varphi} e^{\mathrm{i}\varphi_0\hat{n}}|n\rangle = (2\pi)^{-1}\sum_{n=0}^{\infty} e^{\mathrm{i}n\varphi}[\sum_{m=0}^{\infty}(\mathrm{i}\varphi_0)^m \hat{n}^m|n\rangle/m!]$$

$$= (2\pi)^{-1}\sum_{n=0}^{\infty} e^{\mathrm{i}n\varphi}[\sum_{m=0}^{\infty}(\mathrm{i}\varphi_0 n)^m/m!]|n\rangle = (2\pi)^{-1}\sum_{n=0}^{\infty} e^{\mathrm{i}n\varphi} e^{\mathrm{i}n\varphi_0}|n\rangle$$

$$= (2\pi)^{-1}\sum_{n=0}^{\infty} e^{\mathrm{i}n(\varphi+\varphi_0)}|n\rangle = |\varphi + \varphi_0\rangle, \qquad (0 \le \varphi + \varphi_0 < 2\pi). \tag{38}$$

This is similar to the action of the operator $e^{\mathrm{i}(p_0/\hbar)\hat{x}}$ on the eigenstate $\psi(x) = e^{\mathrm{i}(p/\hbar)x}$ of the momentum operator $\hat{p}_x = -\mathrm{i}\hbar\partial/\partial x$ for a free particle, which yields the shifted eigenstate $e^{\mathrm{i}(p+p_0)x/\hbar}$. Another example is

$$\psi(x + x_0) = \psi(x) + \sum_{n=1}^{\infty} [\partial^n\psi(x)/\partial x^n] x_0^n/n! = |\psi\rangle + \sum_{n=1}^{\infty}(x_0^n/n!)(\mathrm{i}\hat{p}_x/\hbar)^n|\psi\rangle$$

$$= \sum_{n=0}^{\infty}[(\mathrm{i}x_0\hat{p}_x/\hbar)^n/n!]|\psi\rangle = e^{\mathrm{i}x_0\hat{p}_x/\hbar}|\psi\rangle. \tag{39}$$

Finally, we consider the action of the $E$-field operator on the phase eigenstate $|\varphi\rangle$; that is,

$$\langle\varphi|\hat{\boldsymbol{E}}(\boldsymbol{r},t)|\varphi\rangle = \mathrm{i}\sqrt{\hbar\omega/(2\varepsilon_0 V)}\,\langle\varphi|[\hat{\boldsymbol{e}} e^{\mathrm{i}(\boldsymbol{k}\cdot\boldsymbol{r} - \omega t)}\hat{a} - \hat{\boldsymbol{e}}^* e^{-\mathrm{i}(\boldsymbol{k}\cdot\boldsymbol{r} - \omega t)}\hat{a}^\dagger]|\varphi\rangle$$

$$= (\mathrm{i}/2\pi)\sqrt{\hbar\omega/(2\varepsilon_0 V)}\sum_{m=0}^{\infty} e^{-\mathrm{i}m\varphi}\langle m|[\hat{\boldsymbol{e}} e^{\mathrm{i}(\boldsymbol{k}\cdot\boldsymbol{r} - \omega t)}\hat{a} - \hat{\boldsymbol{e}}^* e^{-\mathrm{i}(\boldsymbol{k}\cdot\boldsymbol{r} - \omega t)}\hat{a}^\dagger]\sum_{n=0}^{\infty} e^{\mathrm{i}n\varphi}|n\rangle$$

$$= (\mathrm{i}/2\pi)\sqrt{\hbar\omega/(2\varepsilon_0 V)}\sum_{m=0}^{\infty}\sum_{n=0}^{\infty} e^{\mathrm{i}(n-m)\varphi}[\hat{\boldsymbol{e}} e^{\mathrm{i}(\boldsymbol{k}\cdot\boldsymbol{r} - \omega t)}\langle m|\hat{a}|n\rangle - \hat{\boldsymbol{e}}^* e^{-\mathrm{i}(\boldsymbol{k}\cdot\boldsymbol{r} - \omega t)}\langle m|\hat{a}^\dagger|n\rangle]$$

---

[**] The ½ coefficient of $\delta(\varphi - \varphi')$ should not be concerning. Recall that the momentum eigenstates $\exp(\mathrm{i}px/\hbar)$ of a free particle are often treated as orthonormalized, despite the fact that $\langle p_2|p_1\rangle = \int_{-\infty}^{\infty} e^{\mathrm{i}(p_1-p_2)x/\hbar}\mathrm{d}x = 2\pi\hbar\delta(p_1 - p_2)$.

$$= (\mathrm{i}/2\pi)\sqrt{\hbar\omega/(2\varepsilon_0 V)} \sum_{m=0}^{\infty} \left[\hat{e}e^{\mathrm{i}(\boldsymbol{k}\cdot\boldsymbol{r}-\omega t+\varphi)}\sqrt{m+1} - \hat{e}^* e^{-\mathrm{i}(\boldsymbol{k}\cdot\boldsymbol{r}-\omega t+\varphi)}\sqrt{m}\,\right]$$

$$= (\mathrm{i}/2\pi)\sqrt{\hbar\omega/(2\varepsilon_0 V)} \left(\sum_{m=1}^{\infty}\sqrt{m}\,\right)\left[\hat{e}e^{\mathrm{i}(\boldsymbol{k}\cdot\boldsymbol{r}-\omega t+\varphi)} - \hat{e}^* e^{-\mathrm{i}(\boldsymbol{k}\cdot\boldsymbol{r}-\omega t+\varphi)}\right]$$

$$= -\sqrt{\hbar\omega/(2\pi^2\varepsilon_0 V)} \left(\sum_{m=1}^{\infty}\sqrt{m}\,\right)[\boldsymbol{e}'\sin(\boldsymbol{k}\cdot\boldsymbol{r}-\omega t+\varphi) + \boldsymbol{e}''\cos(\boldsymbol{k}\cdot\boldsymbol{r}-\omega t+\varphi)]. \qquad (40)$$

Note that the expected value of the $E$-field amplitude is in fact infinite, since $\sum_{m=1}^{\infty}\sqrt{m}$ diverges.

One may express arbitrary states of a single-mode field as superpositions of phase eigenstates. This requires writing each number state $|n\rangle$ as a superposition of the eigenstates $|\varphi\rangle$, as follows:

$$|\psi\rangle = \sum_{n=0}^{\infty} c_n |n\rangle \stackrel{\text{see Eq.(32)}}{=} \sum_{n=0}^{\infty} c_n \int_{\varphi=0}^{2\pi} (e^{-\mathrm{i}n\varphi}/\sqrt{2\pi})|\varphi\rangle \mathrm{d}\varphi = (2\pi)^{-\frac{1}{2}} \int_{\varphi=0}^{2\pi} \left(\sum_{n=0}^{\infty} c_n e^{-\mathrm{i}n\varphi}\right)|\varphi\rangle \mathrm{d}\varphi. \qquad (41)$$

Clearly, the amplitude of the state $|\varphi\rangle$ appearing in the integrand of Eq.(41) is a periodic function of $\varphi$ whose Fourier series consists of the amplitudes $c_n$ of the various number states $|n\rangle$. It is not difficult to verify that the probability density $p_\psi(\varphi) = (2\pi)^{-1} \sum_{m=0}^{\infty}\sum_{n=0}^{\infty} c_m c_n^* e^{\mathrm{i}(n-m)\varphi}$ of the state $|\psi\rangle$ being in the phase eigenstate $|\varphi\rangle$ integrates to 1.0 over any interval of length $2\pi$.

**9. Does a phase operator exist?** Considering that a hypothetical phase operator $\hat{\varphi}$ acting on the state $|\psi\rangle$ of Eq.(41) should multiply each eigenstate $|\varphi\rangle$ by its eigenvalue $\varphi$, we must have

$$\hat{\varphi}|\psi\rangle = (2\pi)^{-\frac{1}{2}} \int_{\varphi=0}^{2\pi} \left(\sum_{n=0}^{\infty} c_n e^{-\mathrm{i}n\varphi}\right)\varphi|\varphi\rangle \mathrm{d}\varphi = (2\pi)^{-1} \int_{\varphi=0}^{2\pi}\left(\sum_{n=0}^{\infty} c_n e^{-\mathrm{i}n\varphi}\right)\varphi \sum_{m=0}^{\infty} e^{\mathrm{i}m\varphi}|m\rangle\,\mathrm{d}\varphi$$

$$= (2\pi)^{-1} \sum_{m=0}^{\infty}\sum_{n=0}^{\infty} c_n \int_{\varphi=0}^{2\pi}\varphi e^{\mathrm{i}(m-n)\varphi}\mathrm{d}\varphi\,|m\rangle \stackrel{\text{integration by parts}}{=} \sum_{m=0}^{\infty}[\pi c_m + \mathrm{i}\sum_{\substack{n=0\\n\neq m}}^{\infty} c_n/(n-m)]|m\rangle. \qquad (42)$$

The above equation shows the hypothetical $\hat{\varphi}$ acting properly (i.e., as a Hermitian operator) on the state $|\psi\rangle$. Unfortunately, in the special case of $|\psi\rangle = |\varphi\rangle = (2\pi)^{-\frac{1}{2}}\sum_{n=0}^{\infty} e^{\mathrm{i}n\varphi}|n\rangle$, the operator $\hat{\varphi}$ does *not* yield the expected result, namely, $\hat{\varphi}|\varphi\rangle = \varphi|\varphi\rangle$. This is because the amplitude of $|m\rangle$ in this case turns out to be

$$\pi c_m + \mathrm{i}\sum_{\substack{n=0\\n\neq m}}^{\infty} c_n/(n-m) = (2\pi)^{-\frac{1}{2}}\left[\pi e^{\mathrm{i}m\varphi} + \mathrm{i}\sum_{\substack{n=0\\n\neq m}}^{\infty} e^{\mathrm{i}n\varphi}/(n-m)\right]$$

$$= (2\pi)^{-\frac{1}{2}} e^{\mathrm{i}m\varphi}\left[\pi + \mathrm{i}\sum_{\substack{n=-m\\n\neq 0}}^{\infty} e^{\mathrm{i}n\varphi}/n\right]$$

$$\stackrel{\text{Gradshteyn \& Ryzhik}^{12}\ \mathbf{1.513\text{-}4}}{=} (2\pi)^{-\frac{1}{2}} e^{\mathrm{i}m\varphi}\left[\pi + \mathrm{i}\left(\sum_{\substack{n=-m\\n\neq 0}}^{0} e^{\mathrm{i}n\varphi}/n\right) - \mathrm{i}\ln(1-e^{\mathrm{i}\varphi})\right]. \qquad (43)$$

If, in the definition of the eigenstates $|\varphi\rangle$ in accordance with Eq.(30), the range of $\varphi$ is taken to be $(-\pi, \pi]$, Eq.(42) would become

$$\hat{\varphi}|\psi\rangle = \sum_{m=0}^{\infty}[\mathrm{i}\sum_{\substack{n=0\\n\neq m}}^{\infty}(-1)^{n-m} c_n/(n-m)]|m\rangle. \qquad (44)$$

The corresponding amplitude of the state $|m\rangle$ of $\hat{\varphi}|\varphi\rangle$ now turns out to be

$$\mathrm{i}(2\pi)^{-\frac{1}{2}}\sum_{\substack{n=0\\n\neq m}}^{\infty}(-1)^{n-m} e^{\mathrm{i}n\varphi}/(n-m) = \mathrm{i}(2\pi)^{-\frac{1}{2}} e^{\mathrm{i}m\varphi}\sum_{\substack{n=-m\\n\neq 0}}^{\infty}(-1)^n e^{\mathrm{i}n\varphi}/n$$

$$\stackrel{\text{G\&R}^{12}\ \mathbf{1.511}}{=} (2\pi)^{-\frac{1}{2}} e^{\mathrm{i}m\varphi}\left\{\mathrm{i}[\sum_{\substack{n=-m\\n\neq 0}}^{0}(-1)^n e^{\mathrm{i}n\varphi}/n] - \mathrm{i}\ln(1+e^{\mathrm{i}\varphi})\right\}. \qquad (45)$$

Neither Eq.(43) nor Eq.(45) produces the expected amplitude, namely, $(2\pi)^{-\frac{1}{2}}\varphi e^{im\varphi}$, for the state $|m\rangle$ appearing within $\hat{\varphi}|\varphi\rangle$.[††]

Our hypothetical phase operator $\hat{\varphi}$ has the interesting property that $e^{\pm i\hat{\varphi}} = \sum_{m=0}^{\infty}(\pm i\hat{\varphi})^m/m!$ acting on the number state $|n\rangle = (2\pi)^{-\frac{1}{2}}\int_{\varphi=0}^{2\pi}e^{-in\varphi}|\varphi\rangle d\varphi$ yields $|n \mp 1\rangle$. This is the counterpart of Eq.(38) and shows that $\hat{n}$ and $\hat{\varphi}$ behave in ways that are expected from a pair of conjugate operators — see Eq.(39) for the corresponding relation between position and momentum operators. However, proceeding to evaluate

[For negative $n$, Eq.(32) yields $|n\rangle = 0$.]

$$e^{\pm i\hat{\varphi}}|\varphi\rangle = (2\pi)^{-\frac{1}{2}}e^{\pm i\hat{\varphi}}\sum_{n=0}^{\infty}e^{in\varphi}|n\rangle = (2\pi)^{-\frac{1}{2}}\sum_{n=0}^{\infty}e^{in\varphi}|n \mp 1\rangle, \tag{46}$$

we find the correct answer, namely, $e^{i\varphi}|\varphi\rangle$, for $e^{i\hat{\varphi}}|\varphi\rangle$, but an *incorrect* answer, $e^{-i\varphi}(|\varphi\rangle - |0\rangle)$, for $e^{-i\hat{\varphi}}|\varphi\rangle$. Appendix A provides an alternative treatment of the shifting property of $e^{\pm i\hat{\varphi}}$, which, once again, exposes the inconsistencies in the notional viability of a phase operator.

Another harbinger of potential difficulties with the hypothetical operator $\hat{\varphi}$ of Eq.(42) is that its commutator with the number operator $\hat{n}$ does *not* equal i, as one would expect from a pair of conjugate operators. This is because

$$[\hat{n},\hat{\varphi}]|\psi\rangle = \hat{n}\hat{\varphi}|\psi\rangle - \hat{\varphi}\hat{n}|\psi\rangle$$
$$= \sum_{m=0}^{\infty}[\pi c_m + i\sum_{\substack{n=0\\n\neq m}}^{\infty}c_n/(n-m)]m|m\rangle - \sum_{m=0}^{\infty}[\pi mc_m + i\sum_{\substack{n=0\\n\neq m}}^{\infty}nc_n/(n-m)]|m\rangle$$
$$= -i\sum_{m=0}^{\infty}(\sum_{\substack{n=0\\n\neq m}}^{\infty}c_n)|m\rangle = -i(\sum_{n=0}^{\infty}c_n)(\sum_{m=0}^{\infty}|m\rangle) + i\sum_{m=0}^{\infty}c_m|m\rangle. \tag{47}$$

While the second term on the right-hand side of Eq.(47) is the expected state $i|\psi\rangle$, the unavoidable appearance of the first term in this expression remains problematic.

All in all, the insurmountable difficulties facing our hypothetical phase operator $\hat{\varphi}$ cannot be swept aside. Given that the definition of $\hat{\varphi}$ via Eq.(42) (or Eq.(44)) is a direct consequence of our adopted phase eigenstate $|\varphi\rangle$ in Eq.(30), one must remain mindful of the inaccuracies or unphysical results that might emerge from blind applications of our postulated phase eigenstate $|\varphi\rangle$.

As a substitute for $\hat{\varphi}$, one could consider the operators $½(\hat{a}^\dagger + \hat{a})$ and $½i(\hat{a}^\dagger - \hat{a})$ to estimate the cosine and sine of the phase angle $\varphi$. Considering that $\hat{a}$ and $\hat{a}^\dagger$ transform $|n\rangle$ to $\sqrt{n}|n-1\rangle$ and $\sqrt{n+1}|n+1\rangle$, respectively, further normalization by $\langle n\rangle^{\frac{1}{2}}$ would be needed to adjust the scale of the resulting estimates of $\cos\varphi$ and $\sin\varphi$.[11] In this way, the expected values of $\cos\varphi$, $\sin\varphi$, $\cos^2\varphi$, $\sin^2\varphi$, etc., can be computed for various states of the single-mode $(\omega, \boldsymbol{k}, \hat{\boldsymbol{e}})$ field. For instance, in the simple case of the number state $|n\rangle$, whose average photon number is $\langle n\rangle = \langle n|\hat{n}|n\rangle = n$, we find

$$\langle\cos\varphi\rangle = \langle n|\frac{(\hat{a}^\dagger+\hat{a})}{2\langle n\rangle^{\frac{1}{2}}}|n\rangle = 0. \tag{48}$$

$$\langle\sin\varphi\rangle = \langle n|\frac{i(\hat{a}^\dagger-\hat{a})}{2\langle n\rangle^{\frac{1}{2}}}|n\rangle = 0. \tag{49}$$

$$\langle\cos^2\varphi\rangle = \langle n|\frac{(\hat{a}^\dagger+\hat{a})^2}{4\langle n\rangle}|n\rangle = \langle n|\frac{\hat{a}^\dagger\hat{a}^\dagger+\hat{a}\hat{a}+2\hat{a}^\dagger\hat{a}+1}{4n}|n\rangle = \frac{n+½}{2n}. \tag{50}$$

$$\langle\sin^2\varphi\rangle = \langle n|\frac{i^2(\hat{a}^\dagger-\hat{a})^2}{4\langle n\rangle}|n\rangle = \langle n|\frac{-\hat{a}^\dagger\hat{a}^\dagger-\hat{a}\hat{a}+2\hat{a}^\dagger\hat{a}+1}{4n}|n\rangle = \frac{n+½}{2n}. \tag{51}$$

---

[††] The sums in Eqs.(43) and (45) can be written as $-\sum_{n=1}^{m}[(\pm 1)^n e^{-in\varphi}/n]$. If it were allowed to set $m = \infty$, these sums would converge to $\ln(1 \mp e^{-i\varphi})$, in which case the amplitude of $|m\rangle$ in both Eqs.(43) and (45) would be $(2\pi)^{-\frac{1}{2}}\varphi e^{im\varphi}$.

These results are consistent with one's understanding that the number state has a uniform phase distribution over the $[0, 2\pi)$ interval and that, therefore, its average values of $\cos\varphi$ and $\sin\varphi$ should be zero, while its average values of $\cos^2\varphi$ and $\sin^2\varphi$ should be at or around ½. In the case of Glauber's coherent state $|\gamma\rangle$, where $\gamma = |\gamma|e^{i\theta}$ and $\langle n\rangle = |\gamma|^2$, we have

$$\langle\cos\varphi\rangle = \left[e^{-\frac{1}{2}|\gamma|^2}\sum_{n=0}^{\infty}\langle n|(\gamma^{*n}/\sqrt{n!})\right]\frac{(\hat{a}^\dagger+\hat{a})}{2\langle n\rangle^{\frac{1}{2}}}\left[e^{-\frac{1}{2}|\gamma|^2}\sum_{n=0}^{\infty}(\gamma^n/\sqrt{n!})|n\rangle\right] = \frac{\gamma^*+\gamma}{2|\gamma|} = \cos\theta. \quad (52)$$

$$\langle\sin\varphi\rangle = \langle\gamma|\frac{i(\hat{a}^\dagger-\hat{a})}{2\langle n\rangle^{\frac{1}{2}}}|\gamma\rangle = \frac{i(\gamma^*-\gamma)}{2|\gamma|} = \sin\theta. \quad (53)$$

$$\langle\cos^2\varphi\rangle = \langle\gamma|\frac{\hat{a}^\dagger\hat{a}^\dagger+\hat{a}\hat{a}+2\hat{a}^\dagger\hat{a}+1}{4|\gamma|^2}|\gamma\rangle = \frac{\gamma^{*2}+\gamma^2+2|\gamma|^2+1}{4|\gamma|^2} = \frac{|\gamma|^2\cos 2\theta+|\gamma|^2+\frac{1}{2}}{2|\gamma|^2} = \cos^2\theta + \frac{1}{4|\gamma|^2}. \quad (54)$$

$$\langle\sin^2\varphi\rangle = \langle\gamma|\frac{-\hat{a}^\dagger\hat{a}^\dagger-\hat{a}\hat{a}+2\hat{a}^\dagger\hat{a}+1}{4|\gamma|^2}|\gamma\rangle = \frac{-\gamma^{*2}-\gamma^2+2|\gamma|^2+1}{4|\gamma|^2} = \frac{-|\gamma|^2\cos 2\theta+|\gamma|^2+\frac{1}{2}}{2|\gamma|^2} = \sin^2\theta + \frac{1}{4|\gamma|^2}. \quad (55)$$

The variances of both $\cos\varphi$ and $\sin\varphi$ for the coherent state are seen to be $(4|\gamma|^2)^{-1}$, which approaches zero with an increasing average photon number $\langle n\rangle = |\gamma|^2$. Given a sufficiently large $\langle n\rangle$, a coherent single-mode beam of light can be said to have a well-defined phase $\varphi = \theta$.

In the case of the phase eigenstate $|\varphi\rangle$ of Eq.(30), whose average photon number is $\langle n\rangle = \langle\varphi|\hat{n}|\varphi\rangle = (2\pi)^{-1}\sum_{n=0}^{\infty} n = \infty$, we must resort to an appropriate limit argument to handle the infinities that inevitably arise in these types of calculation.[6] Suppose the allowed number states in $|\varphi\rangle$ are truncated at $n = n_{max}$, and then the phase eigenstate is redefined (in the limit of $n_{max} \to \infty$) as

$$|\varphi\rangle = [2\pi(n_{max} + 1)]^{-\frac{1}{2}}\sum_{n=0}^{n_{max}} e^{in\varphi}|n\rangle. \quad (56)$$

We will find

$$\langle n\rangle = \langle\varphi|\hat{n}|\varphi\rangle = [2\pi(n_{max} + 1)]^{-1}\sum_{n=0}^{n_{max}} n = (4\pi)^{-1}n_{max}. \quad (57)$$

Straightforward evaluation of the expected value of $\cos\varphi$ now yields

$$\langle\cos\varphi\rangle = [2\pi(n_{max} + 1)]^{-1}\left(\sum_{n=0}^{n_{max}}\langle n|e^{-in\varphi}\right)\frac{(\hat{a}^\dagger+\hat{a})}{2\langle n\rangle^{\frac{1}{2}}}\left(\sum_{n=0}^{n_{max}} e^{in\varphi}|n\rangle\right)$$

$$= \frac{1}{2\sqrt{\pi}(n_{max}+1)n_{max}^{\frac{1}{2}}}\left(\sum_{n=0}^{n_{max}}\langle n|e^{-in\varphi}\right)\left(\sum_{n=0}^{n_{max}} e^{in\varphi}\sqrt{n+1}|n+1\rangle + \sum_{n=0}^{n_{max}} e^{in\varphi}\sqrt{n}|n-1\rangle\right)$$

$$= \frac{1}{2\sqrt{\pi}(n_{max}+1)n_{max}^{\frac{1}{2}}}\left(\sum_{n=0}^{n_{max}}\langle n|e^{-in\varphi}\right)\left[\sum_{n=1}^{n_{max}+1} e^{i(n-1)\varphi}\sqrt{n}|n\rangle + \sum_{n=0}^{n_{max}-1} e^{i(n+1)\varphi}\sqrt{n+1}|n\rangle\right]$$

$$= \frac{1}{2\sqrt{\pi}(n_{max}+1)n_{max}^{\frac{1}{2}}}\left(e^{-i\varphi}\sum_{n=1}^{n_{max}}\sqrt{n} + e^{i\varphi}\sum_{n=0}^{n_{max}-1}\sqrt{n+1}\right) = \frac{\sum_{n=1}^{n_{max}}\sqrt{n}}{\sqrt{\pi}(n_{max}+1)n_{max}^{\frac{1}{2}}}\cos\varphi. \quad (58)$$

A good approximation to $\sum_{n=1}^{n_{max}}\sqrt{n}$ is given by $\int_0^{n_{max}}\sqrt{x}dx = 2n_{max}^{3/2}/3$, which essentially cancels out the denominator appearing on the right-hand side of Eq.(58), yielding $\langle\cos\varphi\rangle \cong \cos\varphi$ in the limit when $n_{max} \to \infty$. As for the expected value of $\cos^2\varphi$, we find

$$\langle\cos^2\varphi\rangle = \frac{1}{2n_{max}(n_{max}+1)}\left(\sum_{n=0}^{n_{max}}\langle n|e^{-in\varphi}\right)(\hat{a}^\dagger\hat{a}^\dagger + \hat{a}\hat{a} + 2\hat{a}^\dagger\hat{a} + 1)\left(\sum_{n=0}^{n_{max}} e^{in\varphi}|n\rangle\right)$$

$$= \frac{1}{2n_{max}(n_{max}+1)}\left(\sum_{n=0}^{n_{max}}\langle n|e^{-in\varphi}\right)\left[\sum_{n=0}^{n_{max}} e^{in\varphi}\sqrt{(n+1)(n+2)}|n+2\rangle\right.$$

$$\left. + \sum_{n=0}^{n_{max}} e^{in\varphi}\sqrt{n(n-1)}|n-2\rangle + \sum_{n=0}^{n_{max}} e^{in\varphi}(2n+1)|n\rangle\right]$$

$$= \frac{1}{2n_{\max}(n_{\max}+1)} \left( \sum_{n=0}^{n_{\max}} \langle n|e^{-in\varphi} \right) \left[ \sum_{n=2}^{n_{\max}+2} e^{i(n-2)\varphi} \sqrt{n(n-1)} |n\rangle \right.$$

$$\left. + \sum_{n=0}^{n_{\max}-2} e^{i(n+2)\varphi} \sqrt{(n+2)(n+1)} |n\rangle + \sum_{n=0}^{n_{\max}} e^{in\varphi}(2n+1)|n\rangle \right]$$

$$= \frac{1}{2n_{\max}(n_{\max}+1)} \left[ e^{-i2\varphi} \sum_{n=2}^{n_{\max}} \sqrt{n(n-1)} + e^{i2\varphi} \sum_{n=0}^{n_{\max}-2} \sqrt{(n+2)(n+1)} \right.$$

$$\left. + \sum_{n=0}^{n_{\max}} (2n+1) \right]$$

$$= \frac{\sum_{n=2}^{n_{\max}} \sqrt{n(n-1)}}{n_{\max}(n_{\max}+1)} \left[ (2\cos^2\varphi - 1) + \frac{\sum_{n=0}^{n_{\max}}(n+\frac{1}{2})}{\sum_{n=2}^{n_{\max}} \sqrt{n(n-1)}} \right] \cong \cos^2\varphi \qquad (n_{\max} \to \infty). \tag{59}$$

Similar calculations show that, for the phase eigenstate $|\varphi\rangle$ defined via Eq.(56) in the limit of $n_{\max} \to \infty$, the estimated values of $\langle \sin\varphi \rangle$ and $\langle \sin^2\varphi \rangle$ are very close to $\sin\varphi$ and $\sin^2\varphi$, respectively. Given the reasonableness of the estimated values obtained in the above examples, the operators for $\cos\varphi$ and $\sin\varphi$ are commonly taken as $(\hat{a}^\dagger + \hat{a})/2\langle n\rangle^{1/2}$ and $i(\hat{a}^\dagger - \hat{a})/2\langle n\rangle^{1/2}$, with $\langle n\rangle$ being the expected value of the photon number for the single-mode state under consideration.[11]

**10. Quantum-optical properties of lossless beam-splitters**. The fundamental properties of lossless beam-splitters having Fresnel reflection and transmission coefficients $\rho$ and $\tau$ (with $|\rho|^2 + |\tau|^2 = 1$) have been studied extensively in the literature.[6] The following relations among the annihilation and creation operators at the input ports 1 and 2, and those at the exit ports 3 and 4, are well established:

$$\hat{a}_3 = \rho\hat{a}_1 + \tau\hat{a}_2, \qquad \hat{a}_4 = \tau\hat{a}_1 + \rho\hat{a}_2. \tag{60}$$

$$\hat{a}_3^\dagger = \rho^*\hat{a}_1^\dagger + \tau^*\hat{a}_2^\dagger, \qquad \hat{a}_4^\dagger = \tau^*\hat{a}_1^\dagger + \rho^*\hat{a}_2^\dagger. \tag{61}$$

$$\hat{a}_1^\dagger = \rho\hat{a}_3^\dagger + \tau\hat{a}_4^\dagger, \qquad \hat{a}_2^\dagger = \tau\hat{a}_3^\dagger + \rho\hat{a}_4^\dagger. \tag{62}$$

The splitter is assumed to have front-to-back symmetry, with the same $(\rho, \tau)$ for both input ports 1 and 2. In this case, $\varphi_\rho - \varphi_\tau = \pm\pi/2$.

Also established is the entangled joint photon-number distribution at the exit ports upon simultaneous arrival of the joint number-state $|n_1\rangle_1|n_2\rangle_2$ at the input ports, namely,[13]

$$|n_1\rangle_1|n_2\rangle_2 = \frac{(\hat{a}_1^\dagger)^{n_1}(\hat{a}_2^\dagger)^{n_2}}{\sqrt{n_1!n_2!}} |0\rangle_1|0\rangle_2$$

$$= \sum_{m_1=0}^{n_1} \sum_{m_2=0}^{n_2} \frac{\sqrt{n_1!n_2!(m_1+m_2)!(n_1+n_2-m_1-m_2)!}}{m_1!m_2!(n_1-m_1)!(n_2-m_2)!} \rho^{(n_2+m_1-m_2)} \tau^{(n_1-m_1+m_2)} |m_1+m_2\rangle_3 |n_1+n_2-m_1-m_2\rangle_4.$$

$$\tag{63}^{\ddagger\ddagger}$$

In the special case when $n_1 = n$ and $n_2 = 0$, Eq.(63) reduces to

$$|n\rangle_1|0\rangle_2 = \sum_{m=0}^{n} \binom{n}{m}^{1/2} \rho^m \tau^{n-m} |m\rangle_3 |n-m\rangle_4. \tag{64}$$

The splitter need *not* be symmetric in this case, since $(\rho', \tau')$ are not being used.

It is now easy to see how a Glauber coherent state $|\gamma\rangle$ entering through port 1 of the splitter divides itself between the exit ports 3 and 4; that is,

$$|\gamma\rangle_1|0\rangle_2 = e^{-\frac{1}{2}|\gamma|^2} \sum_{n=0}^{\infty} (\gamma^n/\sqrt{n!}) |n\rangle_1|0\rangle_2$$

$$= e^{-\frac{1}{2}|\gamma|^2} \sum_{n=0}^{\infty} (\gamma^n/\sqrt{n!}) \sum_{m=0}^{n} \binom{n}{m}^{1/2} \rho^m \tau^{n-m} |m\rangle_3 |n-m\rangle_4$$

---

‡‡ For a lossless but asymmetric beam-splitter, the Fresnel reflection and transmission coefficients at ports 1 and 2 will be $(\rho, \tau) = (|\rho|e^{i\varphi_\rho}, |\tau|e^{i\varphi_\tau})$ and $(\rho', \tau') = (|\rho'|e^{i\varphi'_\rho}, |\tau'|e^{i\varphi'_\tau})$, respectively, where $|\rho|^2 + |\tau|^2 = |\rho'|^2 + |\tau'|^2 = 1$. One can show that, in general, $|\rho| = |\rho'|$, $|\tau| = |\tau'|$, and $\varphi_\rho + \varphi'_\rho = \varphi_\tau + \varphi'_\tau \pm \pi$.[13] For such asymmetric splitters, Eq.(63) is modified by replacing $\rho^{(n_2+m_1-m_2)}$ with $\rho^{m_1}\rho'^{(n_2-m_2)}$ and $\tau^{(n_1-m_1+m_2)}$ with $\tau^{n_1-m_1}\tau'^{(m_2)}$.

see Fig.3 ⇒ $= e^{-\frac{1}{2}|\gamma|^2} \sum_{m=0}^{\infty} (\gamma^m \rho^m/\sqrt{m!})|m\rangle_3 \sum_{n=m}^{\infty} [\gamma^{n-m} \tau^{n-m}/\sqrt{(n-m)!}\,]|n-m\rangle_4$

$$= e^{-\frac{1}{2}|\rho\gamma|^2} \sum_{m=0}^{\infty} [(\rho\gamma)^m/\sqrt{m!}]|m\rangle_3 \, e^{-\frac{1}{2}|\tau\gamma|^2} \sum_{n=0}^{\infty} [(\tau\gamma)^n/\sqrt{n!}\,]|n\rangle_4 = |\rho\gamma\rangle_3|\tau\gamma\rangle_4. \quad (65)$$

In accordance with Eq.(65), the emergent beams in both exit ports are seen to be coherent, one with the parameter $\rho\gamma$, the other with the parameter $\tau\gamma$. Moreover, the fact that the emergent state is factorized into two pure states indicates that the beams emerging from ports 3 and 4 are *not* entangled. (A similar result can be obtained when two single-mode beams—having the same frequency $\omega$ and the same polarization $\hat{e}$—one in the coherent state $|\gamma_1\rangle$, the other in the coherent state $|\gamma_2\rangle$, enter through ports 1 and 2 of a lossless beam-splitter.) Appendix B shows that the beams emerging from ports 3 and 4 of a lossless beam-splitter will be entangled when the beam entering port 1 is in a pure number-state $|n\rangle$—while port 2 stays dark; that is, in the vacuum state $|0\rangle$.

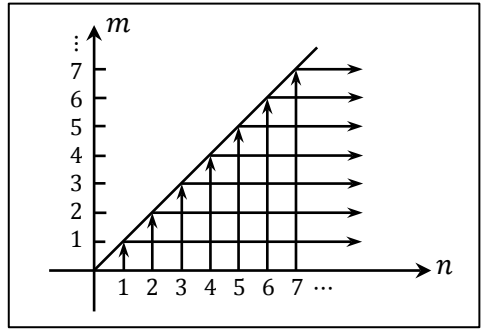

**Fig.3**. The double sum over $m$ and $n$ in Eq.(65) initially ranges from $m = 0$ to $n$ along the vertical axis, and from $n = 0$ to infinity along the horizontal axis. Subsequently, the terms are rearranged so that the inner sum goes horizontally from $n = m$ to infinity, and the outer sum vertically from $m = 0$ to infinity.

As another example of application of the beam-splitter equations (60)-(64), consider the operator $\hat{n}_3 + \hat{n}_4$ acting on the right-hand side of Eq.(63). The operation extracts from $|m_1 + m_2\rangle_3$ the number of photons in port 3, namely, $m = m_1 + m_2$, and from $|n_1 + n_2 - m_1 - m_2\rangle_4$ the number of photons in port 4, namely, $n_1 + n_2 - m = n_1 + n_2 - m_1 - m_2$, then adds them up to arrive at $n_1 + n_2$ for the total number of photons emerging at the exit ports. Pulling $n_1 + n_2$ out of the double sum is all that is needed now to demonstrate that the superposition state on the right-hand side of Eq.(63) is multiplied as a whole by $n_1 + n_2$, thus confirming the equality of $\hat{n}_3 + \hat{n}_4$ and $\hat{n}_1 + \hat{n}_2$. An alternative way of demonstrating the same is via Eqs.(60) and (61), as follows:

$$\hat{n}_3 + \hat{n}_4 = \hat{a}_3^\dagger \hat{a}_3 + \hat{a}_4^\dagger \hat{a}_4 = (\rho^* \hat{a}_1^\dagger + \tau^* \hat{a}_2^\dagger)(\rho \hat{a}_1 + \tau \hat{a}_2) + (\tau^* \hat{a}_1^\dagger + \rho^* \hat{a}_2^\dagger)(\tau \hat{a}_1 + \rho \hat{a}_2)$$

$$= (|\rho|^2 + |\tau|^2)(\hat{a}_1^\dagger \hat{a}_1 + \hat{a}_2^\dagger \hat{a}_2) + (\rho^* \tau + \tau^* \rho)(\hat{a}_1^\dagger \hat{a}_2 + \hat{a}_2^\dagger \hat{a}_1)$$

$\cos(\varphi_\rho - \varphi_\tau) = 0$ ⇒ $= (\hat{a}_1^\dagger \hat{a}_1 + \hat{a}_2^\dagger \hat{a}_2) + 2|\rho||\tau|\cos(\varphi_\rho - \varphi_\tau)(\hat{a}_1^\dagger \hat{a}_2 + \hat{a}_2^\dagger \hat{a}_1) = \hat{n}_1 + \hat{n}_2.$ (66)

The same result can also be demonstrated for asymmetric (albeit lossless) beam-splitters as well.

**11. The Mach-Zehnder interferometer acting as an adjustable beam-splitter**. An ideal Mach-Zehnder interferometer is nothing more than a lossless beam-splitter with an adjustable parameter $\phi$ (i.e., the phase difference between the two arms of the device) that enables one to control its overall reflection and transmission coefficients; see Fig.4.[2,6] We denote by $(\rho_1, \tau_1)$ and $(\rho_2, \tau_2)$ the Fresnel reflection and transmission coefficients of the first and second beam-splitters within the interferometer. (These coefficients are taken to be the same for both entrance ports of each splitter.)
The effective Fresnel coefficients of the interferometer (i.e., the compound beam-splitter) are

$$\rho = \rho_1\rho_2 + \tau_1\tau_2 e^{i\phi}, \quad (67)$$

$$\tau = \rho_1\tau_2 + \tau_1\rho_2 e^{i\phi}. \quad (68)$$

Assuming a single-mode wavepacket $(\omega, \mathbf{k}, \hat{\mathbf{e}})$ in the number state $|n\rangle$ enters through port 1 of the first splitter, the emergent beams at ports 3 and 4 of the second splitter will, in accordance with Eq.(64), be in a superposition of the joint states $|k\rangle_3 |n-k\rangle_4$, as follows:

$$|n\rangle_1 |0\rangle_2 = \sum_{k=0}^{n} \binom{n}{k}^{1/2} \rho^k \tau^{n-k} |k\rangle_3 |n-k\rangle_4. \quad (69)$$

Substitution for $\rho$ and $\tau$ from Eqs.(67) and (68) into Eq.(69) yields

$$|n\rangle_1 |0\rangle_2 = \sum_{k=0}^{n} \binom{n}{k}^{1/2} (\rho_1\rho_2 + \tau_1\tau_2 e^{i\phi})^k (\rho_1\tau_2 + \tau_1\rho_2 e^{i\phi})^{n-k} |k\rangle_3 |n-k\rangle_4$$

$$= \sum_{k=0}^{n} \binom{n}{k}^{1/2} \sum_{m_1=0}^{k} \binom{k}{m_1} (\rho_1\rho_2)^{m_1} (\tau_1\tau_2 e^{i\phi})^{k-m_1} \sum_{m_2=0}^{n-k} \binom{n-k}{m_2} (\rho_1\tau_2)^{m_2} (\tau_1\rho_2 e^{i\phi})^{n-k-m_2} |k\rangle_3 |n-k\rangle_4$$

$$= \sum_{k=0}^{n} \sum_{m_1=0}^{k} \sum_{m_2=0}^{n-k} \binom{n}{k}^{1/2} \binom{k}{m_1}\binom{n-k}{m_2} \rho_1^{m_1+m_2} (\tau_1 e^{i\phi})^{n-m_1-m_2} \rho_2^{n-k+m_1-m_2} \tau_2^{k-m_1+m_2} |k\rangle_3 |n-k\rangle_4$$

$$= \sum_{k=0}^{n} \sum_{m_1=0}^{k} \sum_{m_2=0}^{n-k} \frac{\sqrt{n!\, k!\, (n-k)!}}{m_1!\,(k-m_1)!\,m_2!\,(n-k-m_2)!} \rho_1^{m_1+m_2} (\tau_1 e^{i\phi})^{n-m_1-m_2} \rho_2^{n-k+m_1-m_2} \tau_2^{k-m_1+m_2} |k\rangle_3 |n-k\rangle_4. \quad (70)$$

The various terms appearing in the above equation can be rearranged if we set $m_1 + m_2 = m$ and proceed to rename two of the remaining indices so that $m_1 = k_1$ and $k - m_1 = k_2$. In this way, $m_2$ will be replaced by $m - m_1 = m - k_1$, and $k$ will become $k_1 + k_2$. The index $m$ now ranges from 0 to $n$ (given that $m_1$ can be anywhere between 0 and $k$, while $m_2$ can be anywhere from 0 to $n - k$). Considering that $m_1$ cannot exceed $m$, the new index $k_1$ should range from 0 to $m$. Finally, since $k = k_1 + k_2$ ranges from 0 to $n$, the new index $k_2$ must be between 0 and $n - m$. All in all, we will have

$$|n\rangle_1|0\rangle_2 = \sum_{m=0}^{n}\sum_{k_1=0}^{m}\sum_{k_2=0}^{n-m} \frac{\sqrt{n!\,(k_1+k_2)!\,(n-k_1-k_2)!}}{k_1!\,k_2!\,(m-k_1)!\,(n-m-k_2)!}\rho_1^m(\tau_1 e^{i\phi})^{n-m}\rho_2^{n-m+k_1-k_2}\tau_2^{m-k_1+k_2}|k_1+k_2\rangle_3|n-k_1-k_2\rangle_4$$

$$= \sum_{m=0}^{n}\binom{n}{m}^{1/2}\rho_1^m(\tau_1 e^{i\phi})^{n-m}\sum_{k_1=0}^{m}\sum_{k_2=0}^{n-m}\frac{\sqrt{m!\,(n-m)!\,(k_1+k_2)!\,(n-k_1-k_2)!}}{k_1!\,k_2!\,(m-k_1)!\,(n-m-k_2)!}$$
$$\times \rho_2^{n-m+k_1-k_2}\tau_2^{m-k_1+k_2}|k_1+k_2\rangle_3|n-k_1-k_2\rangle_4. \quad (71)$$

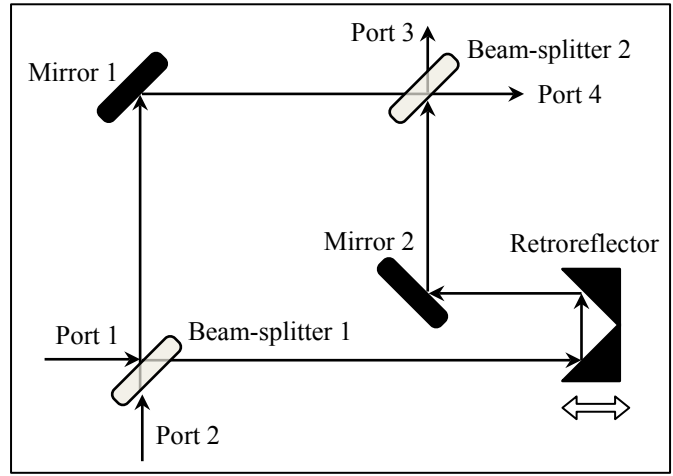

**Fig. 4**. A typical Mach-Zehnder interferometer consists of two beam-splitters, a pair of flat mirrors, and a retroreflector.[2,6] The light can enter through either port 1 or port 2, or simultaneously through both. A fraction of the light then emerges from port 3, while the remaining light leaves through port 4. The optical path-length difference between the two arms of the device introduces a phase shift $\phi$ that can be adjusted by repositioning the retroreflector.

In this way, it is seen that, with a probability amplitude of $\binom{n}{m}^{1/2}\rho_1^m(\tau_1 e^{i\phi})^{n-m}$, the first splitter sends $m$ photons into the entrance port 1, and the remaining $n-m$ photons into the entrance port 2, of the second splitter. The second splitter then divides the arriving photons between its exit ports 3 and 4 in accordance with Eq.(63). Note that the phase-shift factor $e^{i\phi}$ in Eq.(71) is properly combined with the transmission coefficient $\tau_1$ of the first splitter, since it is the beam transmitted through the first splitter whose phase is shifted by $\phi$ prior to arriving at the second splitter.

**12. Degree of first-order coherence**. To simplify the following analysis, we consider a Mach-Zehnder interferometer with a pair of identical 50/50 beam-splitters having $(\rho_1,\tau_1) = (\rho_2,\tau_2) = (1/\sqrt{2}, i/\sqrt{2})$. This results in the compound splitter's reflection and transmission coefficients being $\rho = \sin(\phi/2)\,e^{i(\phi-\pi)/2}$ and $\tau = \cos(\phi/2)\,e^{i(\phi+\pi)/2}$.§§ Let the single-mode wavepacket $(\omega, \boldsymbol{k}, \hat{\boldsymbol{e}})$ entering the input port of the interferometer be in the pure state $|\psi\rangle = \sum_{n=0}^{\infty} c_n|n\rangle$. The output will then be in the (generally entangled) state $\sum_{n=0}^{\infty} c_n \sum_{m=0}^{n}\binom{n}{m}^{1/2}\rho^m\tau^{n-m}|m\rangle_3|n-m\rangle_4$; see Eq.(64). Placing a photodetector in the exit port 3 of the interferometer now yields

$$\langle\psi|\widehat{\boldsymbol{E}}_3^{(-)}\cdot\widehat{\boldsymbol{E}}_3^{(+)}|\psi\rangle = \sum_{n=0}^{\infty} c_n^* \sum_{m=0}^{n}\binom{n}{m}^{1/2}\rho^{*m}\tau^{*(n-m)}\langle m|\langle n-m|\,(-i)\sqrt{\hbar\omega/(2\varepsilon_0 V)}\,\hat{\boldsymbol{e}}^* e^{-i(\boldsymbol{k}\cdot\boldsymbol{r}-\omega t)}\hat{a}_3^\dagger$$
$$\cdot i\sqrt{\hbar\omega/(2\varepsilon_0 V)}\,\hat{\boldsymbol{e}}\,e^{i(\boldsymbol{k}\cdot\boldsymbol{r}-\omega t)}\hat{a}_3 \sum_{n'=0}^{\infty} c_{n'} \sum_{m'=0}^{n'}\binom{n'}{m'}^{1/2}\rho^{m'}\tau^{n'-m'}|m'\rangle|n'-m'\rangle$$
$$= [\hbar\omega/(2\varepsilon_0 V)]\sum_{n=0}^{\infty}|c_n|^2 \sum_{m=0}^{n} m\binom{n}{m}|\rho|^{2m}|\tau|^{2(n-m)} \quad\leftarrow\text{see Appendix B, Eq.(B1)}$$
$$= [\hbar\omega/(2\varepsilon_0 V)]\sum_{n=0}^{\infty}|c_n|^2 n|\rho|^2(|\rho|^2+|\tau|^2)^{n-1}$$
$$= [\langle n\rangle\hbar\omega/(2\varepsilon_0 V)]\sin^2(\phi/2). \quad (72)$$

If $\phi = 0$ (or, in general, any integer-multiple of $2\pi$), the detected signal will be zero; this is due to destructive interference between the two arms of the interferometer—i.e., cancellation of two reflections and two transmissions that occur at the two beam-splitters. In contrast, if $\phi = \pi$ (or any odd-integer-multiple of $\pi$), the expected value of the detected signal will be $\langle n\rangle\hbar\omega/(2\varepsilon_0 V)$; that is, constructive interference causes the input beam in its entirety to emerge from port 3. In the case of $\phi = \pi/2$, one half of the incident optical energy emerges from port 3, while the remaining half exits through port 4.

Repeating the same calculation with the photodetector now placed in port 4, we find the detected signal as $\langle\psi|\widehat{\boldsymbol{E}}_4^{(-)}\cdot\widehat{\boldsymbol{E}}_4^{(+)}|\psi\rangle = [\langle n\rangle\hbar\omega/(2\varepsilon_0 V)]\cos^2(\phi/2)$. The difference between the signals in the exit ports of the Mach-Zehnder interferometer is thus found to be

$$\langle\psi|\widehat{\boldsymbol{E}}_4^{(-)}\cdot\widehat{\boldsymbol{E}}_4^{(+)} - \widehat{\boldsymbol{E}}_3^{(-)}\cdot\widehat{\boldsymbol{E}}_3^{(+)}|\psi\rangle = [\langle n\rangle\hbar\omega/(2\varepsilon_0 V)]\cos\phi = \text{Real}\{[\langle n\rangle\hbar\omega/(2\varepsilon_0 V)]e^{i\omega(t'-t)}\}. \quad (73)$$

---

§§ The 180° phase difference between $\rho$ and $\tau$ of this compound beam-splitter is a consequence of the fact that the splitter lacks front-to-back symmetry; see Appendix C for a detailed analysis of this type of compound beam-splitter.

Here, $\phi = \omega(t' - t)$ is the interferometer's internal phase-shift due to the time-delay between its two arms. It is seen that Eq.(73), which represents the degree of first-order coherence of the incident beam,[2,6] is mathematically equivalent to $\langle\psi|\widehat{\boldsymbol{E}}^{(-)}(t') \cdot \widehat{\boldsymbol{E}}^{(+)}(t)|\psi\rangle$, an expression that can be computed directly, without acknowledging the presence (and the absolute necessity) of the interferometer in the actual measurement system.

**13. Degree of second-order coherence**. To compute the intensity correlation function of a single-mode $(\omega, \boldsymbol{k}, \widehat{\boldsymbol{e}})$ beam in the pure state $|\psi\rangle = \sum_{n=0}^{\infty} c_n |n\rangle$, we place a lossless splitter with Fresnel coefficients $(\rho, \tau)$ in the beam's path, then use a pair of photodetectors in the exit ports 3 and 4 to obtain the correlation between the two detector signals, as follows:

$$\langle 0|_2 \langle\psi|_1 \widehat{E}_3^{(-)} \widehat{E}_4^{(-)} \widehat{E}_4^{(+)} \widehat{E}_3^{(+)} |\psi\rangle_1 |0\rangle_2 \quad \leftarrow \text{assuming linear polarization simplifies the calculation}$$

$$= [\hbar\omega/(2\varepsilon_0 V)]^2 \sum_{n=0}^{\infty} \sum_{m=0}^{n} c_n^* \binom{n}{m}^{1/2} \rho^{*m} \tau^{*(n-m)} \langle n-m|_4 \langle m|_3 \widehat{a}_3^\dagger \widehat{a}_4^\dagger \widehat{a}_4 \widehat{a}_3$$

$$\sum_{n'=0}^{\infty} \sum_{m'=0}^{n'} c_{n'} \binom{n'}{m'}^{1/2} \rho^{m'} \tau^{(n'-m')} |m'\rangle_3 |n'-m'\rangle_4$$

$$= [\hbar\omega/(2\varepsilon_0 V)]^2 \sum_{n=0}^{\infty} \sum_{m=0}^{n} |c_n|^2 m(n-m) \binom{n}{m} |\rho|^{2m} |\tau|^{2(n-m)} \quad \leftarrow \text{see Appendix B}$$

$$= [\hbar\omega/(2\varepsilon_0 V)]^2 \sum_{n=0}^{\infty} |c_n|^2 [n^2 |\rho|^2 - n|\rho|^2 - n(n-1)|\rho|^4]$$

$$= [\hbar\omega/(2\varepsilon_0 V)]^2 |\rho|^2 |\tau|^2 \sum_{n=0}^{\infty} |c_n|^2 n(n-1) = \langle n(n-1)\rangle [\hbar\omega/(2\varepsilon_0 V)]^2 |\rho|^2 |\tau|^2. \qquad (74)$$

The fundamental difference between intensity correlation functions of classical and quantum optics is now manifest in the leading coefficient $\langle n(n-1)\rangle$ appearing on the right-hand side of Eq.(74). Whereas the expected signals in ports 3 and 4 are proportional to $|\rho|^2 \langle n\rangle$ and $|\tau|^2 \langle n\rangle$, the correlation between these signals is proportional to $|\rho|^2 |\tau|^2 \langle n(n-1)\rangle$. This is due to the essential indivisibility of single photons in quantum optics, which dictates that each photon must be detected in either port 3 or port 4, rather than splitting its energy between the two ports, as suggested by classical optics.[2,4-7] Note that, in the special case of a Glauber coherent beam entering the beam-splitter, the identity $\langle n(n-1)\rangle = \langle n^2\rangle - \langle n\rangle = \langle n\rangle^2$ serves as a reminder that not only are the emergent modes in ports 3 and 4 coherent, but also that they are *not* entangled.

**14. Concluding remarks**. Invoking the $E$-field operator of Sec.2, Eq.(1), it is easy to show that a single-mode $(\omega, \boldsymbol{k}, \widehat{\boldsymbol{e}})$ of the EM field residing in a (large) volume $V$ of free space and occupied by $n$ photons in the number-state $|n\rangle$, will have an average $E$-field amplitude $\langle\widehat{\boldsymbol{E}}\rangle = 0$ and variance $\langle\widehat{\boldsymbol{E}} \cdot \widehat{\boldsymbol{E}}\rangle - \langle\widehat{\boldsymbol{E}}\rangle^2 = (n + \tfrac{1}{2})\hbar\omega/(\varepsilon_0 V)$. Given the anticipated energy content $n\hbar\omega$ of this EM wave (which must be equally split between its electric and magnetic fields), the classical $E$-field energy $\tfrac{1}{2}\varepsilon_0 (\boldsymbol{E} \cdot \boldsymbol{E})V$ of the wave comes close to the expected value of its quantum counterpart. The ¼$\hbar\omega$ difference between the classical and quantum values of the $E$-field energy is attributed to vacuum fluctuations, which are present even when the number of photons in the state is $n = 0$. Real-world photodetectors, of course, cannot pick up the vacuum fluctuations, which is why the correct operator for their observed photocurrent is proportional to $\widehat{\boldsymbol{E}}^{(-)} \cdot \widehat{\boldsymbol{E}}^{(+)}$, as elaborated in Sec.2.

It is obvious that putting all the photons of a light beam into a single number-state will be ideal for photon-counting purposes since the corresponding fluctuations, i.e., $\langle\widehat{n}^2\rangle - \langle\widehat{n}\rangle^2$, turn out to be precisely zero. However, it is difficult to monitor or to manipulate the phase of such a beam—which renders interferometric measurements beyond its reach—not only because the expected values $\langle \hat{E}_p \rangle$ and $\langle \hat{E}_q \rangle$ of both $E$-field quadratures now vanish, but also because other relevant operators such as $\hat{a}$, $\hat{a}^\dagger$, $\hat{a}\hat{a}$, $\hat{a}^\dagger\hat{a}^\dagger$, $\hat{a}^\dagger\hat{a}\hat{a}^\dagger$, $\hat{a}^\dagger\hat{a}\hat{a}$, etc., end up having zero expected values when they act on the (one and only) number-state $|n\rangle$.

One way to construct a state $|\psi\rangle$ with a more or less well-defined phase for its $E$-field is to include in $|\psi\rangle$ a reasonably large number of adjacent (or near-adjacent) number-states $|n\rangle$ with probability amplitudes $c_n = |c_n|e^{i\theta_n}$, where the magnitudes $|c_n|$ are distributed fairly uniformly while the phase angles $\theta_n$ exhibit a linear dependence on $n$. The more number-states of this kind one includes in $|\psi\rangle$, the less uncertain the value of its phase becomes, which makes the state suitable for applications involving interferometry; this has been the lesson of Sec.8. Unfortunately, the broad distribution of $|c_n|$ now causes an increase in the photocount fluctuations associated with $|\psi\rangle$, making the corresponding photon-counting signal noisier. The beauty of Glauber's coherent state $|\gamma\rangle$ is that it strikes a fine balance between phase accuracy and photocount fluctuations while keeping the uncertainty product $(\delta E_p)(\delta E_q)$ of the $E$-field quadratures to a minimum.

There exist applications that could benefit from a reduced photon-counting noise at the expense of increased fluctuations elsewhere in the system—or vice-versa. The use of a squeezed state of light in such situations can lead to improved performance.[6,7] Just as a laser can produce a coherent beam of light in a desired $(\omega, \boldsymbol{k}, \hat{\boldsymbol{e}})$ mode with exquisite control over the amplitude and phase of the radiated $E$-field (albeit within the inescapable constraints of the uncertainty principle), so can the tools and techniques of nonlinear optics be corralled to generate squeezed light with specific desirable characteristics.[7] In contrast to coherent light, a squeezed state can be prepared such that its fluctuations are reduced in one quadrature and increased in the other—again subject to the constraints of the uncertainty principle, as discussed in Sec.7.

In recent years, many sensing and control tasks have been brought to operate at or near the quantum limits of accuracy with the aid of classical interferometric arrangements encompassing high-quality optical elements, squeezed as well as coherent light sources, and nearly ideal photo-detectors.[7] It was shown in Sec.11 that a Mach-Zehnder interferometer is essentially an adjustable beam-splitter, whose quantum-optical properties are directly relatable to those of conventional (albeit asymmetric) beam-splitters. A similar argument can be made for Michelson-type interferometers,[2,6] where one or more input beams enter through a 50/50 splitter, bounce off a pair of mirrors at the end of each arm of the instrument, then return through the same ports where the beam(s) had initially entered. A grand example of such systems is the *Laser Interferometer Gravitational-wave Observatory* (LIGO) that, for the first time on 14 September 2015, succeeded in detecting the gravitational wave emitted by a black hole collision event that had occurred nearly a billion light years away from our planet. One can study the subtle and delicate optical aspects of LIGO by extending the methods of this paper to examine the radiation pressure noise acting on the system's primary mirrors, and also to dissect the photodetection noise that accompanies the LIGO interferometer's output signal.[7,9] Specifically, the methods developed in this paper can be used to explain how the injection of a second light beam, this one in the squeezed vacuum state,[6,7] through LIGO's output channel ends up attenuating the photocount fluctuations noise provided that the phase of the injected squeezed light (relative to that of the main laser beam) is properly adjusted.[9] The photon-counting noise, of course, is a major source of contamination of the gravitational-wave signal that is picked up by a photodetector at the exit port of LIGO's Michelson interferometer.

While for pedagogical purposes we restricted the scope of the discussion in this paper to single-mode EM waves that propagate in free space, the results can be extended to multimode waves under more general circumstances. The interested reader should consult the vast literature of the subject.[6,7]

## Appendix A

The shifting property of the hypothetical phase operator $\hat{\varphi}$ described in Sec.9, namely, $e^{\pm i\hat{\varphi}}|n\rangle = |n \mp 1\rangle$, can be related to the annihilation and creation operators in the following way:

$$\hat{a} = e^{i\hat{\varphi}}\hat{n}^{\frac{1}{2}} \quad \text{and} \quad \hat{a}^\dagger = \hat{n}^{\frac{1}{2}}e^{-i\hat{\varphi}}. \tag{A1}$$

Alternative expressions of this shifting property of $e^{\pm i\hat{\varphi}}$ are

$$e^{i\hat{\varphi}} = \sum_{n=0}^{\infty}|n\rangle\langle n+1| \quad \text{and} \quad e^{-i\hat{\varphi}} = \sum_{n=0}^{\infty}|n+1\rangle\langle n|. \tag{A2}$$

If $\hat{\varphi}$ were a proper Hermitian operator, we would have had $e^{i\hat{\varphi}}e^{-i\hat{\varphi}} = e^{-i\hat{\varphi}}e^{i\hat{\varphi}} = \hat{\mathbb{I}}$. However, Eqs.(A2) yield

$$e^{i\hat{\varphi}}e^{-i\hat{\varphi}} = \sum_{n=0}^{\infty}|n\rangle\langle n| = \hat{\mathbb{I}}, \tag{A3}$$

$$e^{-i\hat{\varphi}}e^{i\hat{\varphi}} = \sum_{n=0}^{\infty}|n+1\rangle\langle n+1| = \hat{\mathbb{I}} - |0\rangle\langle 0|. \tag{A4}$$

This shortcoming of $e^{\pm i\hat{\varphi}}$ operators does *not* affect their relation with $\hat{a}$ and $\hat{a}^\dagger$ as expressed by Eqs.(A1), since one still arrives at the expected behavior of $\hat{a}^\dagger\hat{a}$ and $\hat{a}\hat{a}^\dagger$, namely,

$$\hat{a}^\dagger\hat{a} = \hat{n}^{\frac{1}{2}}e^{-i\hat{\varphi}}e^{i\hat{\varphi}}\hat{n}^{\frac{1}{2}} = \hat{n}^{\frac{1}{2}}(\hat{\mathbb{I}} - |0\rangle\langle 0|)\hat{n}^{\frac{1}{2}} = \hat{n}, \tag{A5}$$

$$\hat{a}\hat{a}^\dagger = e^{i\hat{\varphi}}\hat{n}e^{-i\hat{\varphi}} = \sum_{m=0}^{\infty}\sum_{n=0}^{\infty}|m\rangle\langle m+1|(n+1)|n+1\rangle\langle n|$$

$$= \sum_{n=0}^{\infty}n|n\rangle\langle n| + \sum_{n=0}^{\infty}|n\rangle\langle n| = \hat{n} + \hat{\mathbb{I}}. \tag{A6}$$

Nevertheless, the fundamental inconsistencies of the phase operator $\hat{\varphi}$ remain insurmountable, which is why one is driven to use the operators $\frac{1}{2}(\hat{a}^\dagger + \hat{a})/\langle n\rangle^{\frac{1}{2}}$ and $\frac{1}{2}i(\hat{a}^\dagger - \hat{a})/\langle n\rangle^{\frac{1}{2}}$ to estimate the behavior of $\cos\varphi$ and $\sin\varphi$ (albeit to good approximation), as related in the latter half of Sec.9.

## Appendix B

Let a single-mode EM wavepacket in the number-state $|n\rangle$ enter through port 1 of a lossless beam-splitter, whose Fresnel reflection and transmission coefficients are specified as $(\rho, \tau)$; the entrance port 2 is in the dark; that is, in the vacuum state $|0\rangle$. In what follows, we demonstrate the entanglement of the beams that emerge from the splitter's exit ports 3 and 4 by showing that $\langle m(n-m)\rangle_{3,4} \neq \langle m\rangle_3\langle n-m\rangle_4$; here $\langle m\rangle_3$ and $\langle n-m\rangle_4$ are the expected numbers of photons that exit through ports 3 and 4. We begin by proving the following identities:

$$\sum_{m=0}^{n} m\binom{n}{m}x^m y^{n-m} = nx(x+y)^{n-1}. \tag{B1}$$

$$\sum_{m=0}^{n} m^2\binom{n}{m}x^m y^{n-m} = nx(x+y)^{n-1} + n(n-1)x^2(x+y)^{n-2}. \tag{B2}$$

Invoking the binomial expansion $(x+y)^n = \sum_{m=0}^{n}\binom{n}{m}x^m y^{n-m}$, one can write

$$\sum_{m=0}^{n}\binom{n}{m}x^{m+1}y^{n-m} = x\sum_{m=0}^{n}\binom{n}{m}x^m y^{n-m} = x(x+y)^n. \tag{B3}$$

Differentiation with respect to $x$ now yields

$$\frac{d}{dx}\sum_{m=0}^{n}\binom{n}{m}x^{m+1}y^{n-m} = \sum_{m=0}^{n}(m+1)\binom{n}{m}x^m y^{n-m} = (x+y)^n + \sum_{m=0}^{n}m\binom{n}{m}x^m y^{n-m}$$

$$= \tfrac{d}{dx}[x(x+y)^n] = (x+y)^n + nx(x+y)^{n-1}$$

$$\rightarrow \sum_{m=0}^{n} m \binom{n}{m} x^m y^{n-m} = nx(x+y)^{n-1}. \tag{B4}$$

In the special case when $x = y$, the above identity reduces to $\sum_{m=0}^{n} m \binom{n}{m} = 2^{n-1} n$. The identity (B2) is proven in a similar way, as follows:

$$\tfrac{d^2}{dx^2} \sum_{m=0}^{n} \binom{n}{m} x^{m+2} y^{n-m} = \sum_{m=0}^{n}(m+2)(m+1) \binom{n}{m} x^m y^{n-m}$$

$$= 2(x+y)^n + 3 \sum_{m=0}^{n} m \binom{n}{m} x^m y^{n-m} + \sum_{m=0}^{n} m^2 \binom{n}{m} x^m y^{n-m}$$

$$= 2(x+y)^n + 3nx(x+y)^{n-1} + \sum_{m=0}^{n} m^2 \binom{n}{m} x^m y^{n-m}. \tag{B5}$$

$$\tfrac{d^2}{dx^2}[x^2(x+y)^n] = 2(x+y)^n + 4nx(x+y)^{n-1} + n(n-1)x^2(x+y)^{n-2}. \tag{B6}$$

Equating Eqs.(B5) and (B6), we finally arrive at

$$\sum_{m=0}^{n} m^2 \binom{n}{m} x^m y^{n-m} = nx(x+y)^{n-1} + n(n-1)x^2(x+y)^{n-2}. \tag{B7}$$

In the special case when $x = y$, the preceding identity reduces to $\sum_{m=0}^{n} m^2 \binom{n}{m} = 2^{n-2} n(n+1)$.

We are now in a position to evaluate the average (or expected) number of photons that exit through port 3; that is,

$$\langle m \rangle_3 = \langle n-m|_4 \langle m|_3 \sum_{m=0}^{n} \binom{n}{m}^{1/2} \rho^{*m} \tau^{*(n-m)} \, \hat{n}_3 \sum_{m=0}^{n} \binom{n'}{m'}^{1/2} \rho^{m'} \tau^{n'-m'} |m'\rangle_3 |n'-m'\rangle_4$$

$$= \sum_{m=0}^{n} m \binom{n}{m} |\rho|^{2m} |\tau|^{2(n-m)} \overset{\text{see Eq.(B1)}}{=} n|\rho|^2 (|\rho|^2 + |\tau|^2)^{n-1} = n|\rho|^2. \tag{B8}$$

In a similar way, one can obtain $\langle n-m \rangle_4$ using the operator $\hat{n}_4$ or, more straightforwardly, by noting that $\langle n-m \rangle_4 = n - \langle m \rangle_3 = n - n|\rho|^2 = n|\tau|^2$. Finally, the expected value of $m(n-m)$ is obtained by using the operator $\hat{n}_4 \hat{n}_3$ on the emergent superposition state, which results in

$$\langle m(n-m) \rangle_{3,4} = \sum_{m=0}^{n} m(n-m) \binom{n}{m} |\rho|^{2m} |\tau|^{2(n-m)} \overset{\text{see Eqs.(B1) and (B2)}}{=} n^2 |\rho|^2 - n|\rho|^2 - n(n-1)|\rho|^4$$

$$= n(n-1)|\rho|^2 |\tau|^2. \tag{B9}$$

The expected value of $m(n-m)$ is thus seen to be smaller than $\langle m \rangle_3 \langle n-m \rangle_4 = n^2 |\rho|^2 |\tau|^2$, which confirms that the emerging wavepackets in ports 3 and 4 are entangled.

**Appendix C**

The lossless Mach-Zehnder interferometer depicted in Fig.4 consists of a first beam-splitter with Fresnel reflection and transmission coefficients $(\rho_1, \tau_1)$ and a second splitter with the corresponding coefficients $(\rho_2, \tau_2)$. Both beam-splitters are assumed to be symmetric, implying that the phase difference between their respective $\rho$ and $\tau$ is 90°; that is, if $\rho = |\rho|e^{i\varphi}$, then $\tau = |\tau|e^{i(\varphi \pm \frac{1}{2}\pi)}$. The path-length difference between the two arms of the device imparts a phase $\phi$ to the beam in the lower arm relative to that in the upper arm. In what follows, we write expressions for the compound reflection and transmission coefficients when the beam enters through any one of the ports 1, 2, 3, or 4. We proceed to show that, in general, all reflection coefficients have identical magnitudes, as do all transmission coefficients. We also verify that $\rho_{13} = \rho_{31}$, $\rho_{24} = \rho_{42}$, $\tau_{14} = \tau_{41}$, $\tau_{23} = \tau_{32}$, and that $\varphi_{\rho_{13}} + \varphi_{\rho_{24}} = \varphi_{\tau_{14}} + \varphi_{\tau_{23}} \pm \pi$.

The various reflection and transmission coefficients of the interferometer are obtained by combining the coefficients pertaining to passage through the two arms of the device, as follows:

$$\rho_{13} = \rho_1\rho_2 + \tau_1\tau_2 e^{i\phi}; \quad \rho_{24} = \rho_1\rho_2 e^{i\phi} + \tau_1\tau_2; \quad \rho_{31} = \rho_2\rho_1 + \tau_2\tau_1 e^{i\phi}; \quad \rho_{42} = \rho_2\rho_1 e^{i\phi} + \tau_2\tau_1.$$

$$\tau_{14} = \rho_1\tau_2 + \tau_1\rho_2 e^{i\phi}; \quad \tau_{23} = \rho_1\tau_2 e^{i\phi} + \tau_1\rho_2; \quad \tau_{32} = \rho_2\tau_1 + \tau_2\rho_1 e^{i\phi}; \quad \tau_{41} = \rho_2\tau_1 e^{i\phi} + \tau_2\rho_1.$$

Clearly, $\rho_{13} = \rho_{31}$, $\rho_{24} = \rho_{42}$, $\tau_{14} = \tau_{41}$, and $\tau_{23} = \tau_{32}$. Also, recalling that $\varphi_{\tau_1} = \varphi_{\rho_1} \pm \frac{1}{2}\pi$ and $\varphi_{\tau_2} = \varphi_{\rho_2} \pm \frac{1}{2}\pi$, one can write

$$|\rho_{13}|^2 = (\rho_1\rho_2 + \tau_1\tau_2 e^{i\phi})(\rho_1^*\rho_2^* + \tau_1^*\tau_2^* e^{-i\phi}) = |\rho_1\rho_2|^2 + |\tau_1\tau_2|^2 - |\rho_1\rho_2\tau_1\tau_2|(e^{i\phi} + e^{-i\phi}). \quad (C1)$$

$$|\rho_{24}|^2 = (\rho_1\rho_2 e^{i\phi} + \tau_1\tau_2)(\rho_1^*\rho_2^* e^{-i\phi} + \tau_1^*\tau_2^*) = |\rho_1\rho_2|^2 + |\tau_1\tau_2|^2 - |\rho_1\rho_2\tau_1\tau_2|(e^{i\phi} + e^{-i\phi}). \quad (C2)$$

$$|\tau_{14}|^2 = (\rho_1\tau_2 + \tau_1\rho_2 e^{i\phi})(\rho_1^*\tau_2^* + \tau_1^*\rho_2^* e^{-i\phi}) = |\rho_1\tau_2|^2 + |\tau_1\rho_2|^2 + |\rho_1\rho_2\tau_1\tau_2|(e^{i\phi} + e^{-i\phi}). \quad (C3)$$

$$|\tau_{23}|^2 = (\rho_1\tau_2 e^{i\phi} + \tau_1\rho_2)(\rho_1^*\tau_2^* e^{-i\phi} + \tau_1^*\rho_2^*) = |\rho_1\tau_2|^2 + |\tau_1\rho_2|^2 + |\rho_1\rho_2\tau_1\tau_2|(e^{i\phi} + e^{-i\phi}). \quad (C4)$$

It is seen that $|\rho_{13}| = |\rho_{24}|$ and $|\tau_{14}| = |\tau_{23}|$. These and the previously established identities confirm that the magnitudes of all the reflection coefficients are the same, as are the magnitudes of all the transmission coefficients. We also have

$$\rho_{13}\tau_{14}^* = (\rho_1\rho_2 + \tau_1\tau_2 e^{i\phi})(\rho_1^*\tau_2^* + \tau_1^*\rho_2^* e^{-i\phi})$$

$$= |\rho_1|^2\rho_2\tau_2^* + |\tau_1|^2\rho_2^*\tau_2 + |\rho_2|^2\rho_1\tau_1^* e^{-i\phi} + |\tau_2|^2\rho_1^*\tau_1 e^{i\phi}$$

$$= |\rho_2\tau_2|(|\rho_1|^2 e^{\mp i\pi/2} + |\tau_1|^2 e^{\pm i\pi/2}) + |\rho_1\tau_1|[|\rho_2|^2 e^{-i(\phi \pm \frac{1}{2}\pi)} + |\tau_2|^2 e^{i(\phi \pm \frac{1}{2}\pi)}]. \quad (C5)$$

$$\rho_{24}^*\tau_{23} = (\rho_1^*\rho_2^* e^{-i\phi} + \tau_1^*\tau_2^*)(\rho_1\tau_2 e^{i\phi} + \tau_1\rho_2)$$

$$= |\rho_1|^2\rho_2^*\tau_2 + |\tau_1|^2\rho_2\tau_2^* + |\rho_2|^2\rho_1^*\tau_1 e^{-i\phi} + |\tau_2|^2\rho_1\tau_1^* e^{i\phi}$$

$$= |\rho_2\tau_2|(|\rho_1|^2 e^{\pm i\pi/2} + |\tau_1|^2 e^{\mp i\pi/2}) + |\rho_1\tau_1|[|\rho_2|^2 e^{-i(\phi \mp \frac{1}{2}\pi)} + |\tau_2|^2 e^{i(\phi \mp \frac{1}{2}\pi)}]. \quad (C6)$$

Noting that $\rho_{13}\tau_{14}^* + \rho_{24}^*\tau_{23} = 0$, we conclude that $\varphi_{\rho_{13}} - \varphi_{\tau_{14}} = \varphi_{\tau_{23}} - \varphi_{\rho_{24}} \pm \pi$, which may equivalently be written as $\varphi_{\rho_{13}} + \varphi_{\rho_{24}} = \varphi_{\tau_{14}} + \varphi_{\tau_{23}} \pm \pi$.

# References

1. R. P. Feynman, R. B. Leighton, and M. Sands, *The Feynman Lectures on Physics* (Volume III), Addison-Wesley, Massachusetts (1965).
2. L. Mandel and E. Wolf, *Optical Coherence and Quantum Optics*, Cambridge University Press, Cambridge (1995).
3. M. Mansuripur, *Field, Force, Energy and Momenta in Classical Electrodynamics* (Revised Edition), Bentham Science Publishers, Sharjah (2017).
4. R. J. Glauber, "The Quantum Theory of Coherence," *Physical Review* **130**, pp 2529-39 (1963).
5. C. Cohen-Tannoudji, J. Dupont-Roc, and G. Grynberg, *Photons & Atoms: Introduction to Quantum Electrodynamics*, Wiley, New York (1989).
6. R. Loudon, *The Quantum Theory of Light* (3$^{rd}$ edition), Oxford University Press, Oxford, United Kingdom (2000).
7. G. Grynberg, A. Aspect, and C. Fabre, *Introduction to Quantum Optics*, Cambridge University Press, Cambridge, United Kingdom (2010).
8. M. Mansuripur, "Insights into the behavior of certain optical systems gleaned from Feynman's approach to quantum electrodynamics," *Proceedings of SPIE* **12197**, pp 1219703-1:42 (2022); doi: 10.1117/12.2632902.
9. C. M. Caves, "Quantum-mechanical noise in an interferometer," *Physical Review D* **23**, pp 1693-1708 (1981).
10. M. Mansuripur, "Linear and angular momenta of photons in the context of 'which path' experiments of quantum mechanics," *Proceedings of SPIE* **12198**, 1219807-1:12 (2022); doi: 10.1117/12.2632866.
11. S. M. Barnett and D. T. Pegg, "Phase in quantum optics," *J. Phys. A: Math. Gen.* **19**, 3849-3862 (1986).
12. I. S. Gradshteyn & I. M. Ryzhik, *Table of Integrals, Series, and Products* (7$^{th}$ edition), Academic, New York (2007).
13. M. Mansuripur and E. M. Wright, "Fundamental properties of beamsplitters in classical and quantum optics," *Am. J. Phys.* **91**, 298-306 (2023).